\documentclass[12pt]{article}

\usepackage{amsfonts}
\usepackage{amssymb}

\newcommand{\beq}[1]{\begin{equation}\label{#1}}
\newcommand{\eeq}{\end{equation}}
\newcommand{\bear}[1]{\begin{eqnarray}\label{#1}}
\newcommand{\ear}{\end{eqnarray}}
\newcommand{\nn}{\nonumber}

\renewcommand{\theequation}{\arabic{section}.\arabic{equation}}
\catcode`\@=11 \@addtoreset{equation}{section}\catcode`\@=12

\newcommand{\np}{ {\newpage } }

\newcommand{\1}{ {\bf 1 } }

 \newcommand{\R}{ {\mathbb R} }

  \newcommand{\C}{ {\mathbb C} }

 \newcommand{\G}{ {\bf G} }

\newcommand{\p}{\partial}
\newcommand{\btd}{\bigtriangledown}

 \newcommand{\fnm}{\footnotemark}
 \newcommand{\fnt}{\footnotetext}

\begin{document}

\begin{center}    \bf \large

More M-branes  on product of Ricci-flat manifolds

\end{center}

\vspace{0.96truecm}

\begin{center}

 \normalsize\bf V. D. Ivashchuk\fnm[1]\fnt[1]{e-mail:
  ivashchuk@mail.ru}

\vspace{0.3truecm}

 \it Center for Gravitation and Fundamental Metrology,
 VNIIMS, 46 Ozyornaya ul., Moscow 119361, Russia

 \it Institute of Gravitation and Cosmology,
 Peoples' Friendship University of Russia,
 6 Miklukho-Maklaya ul., Moscow 117198, Russia

\end{center}

\begin{abstract}

Partially supersymmetric   intersecting (non-marginal) composite
$M$-brane  solutions defined on the product of Ricci-flat
manifolds $M_{0}  \times M_{1} \times \ldots \times M_{n}$ in $D=
11$ supergravity  are considered and formulae for fractional
numbers of unbroken  supersymmetries are derived for  the
following configurations  of branes:  $M2 \cap M2$, $M2 \cap M5$,
$M5 \cap M5$  and $M2 \cap M2 \cap M2$. Certain examples of
partially supersymmetric configurations  are presented.

\end{abstract}

\np

\section{\bf Introduction}
\setcounter{equation}{0}

In this paper we consider a class of partially supersymmetric
solutions in 11-dimensional supergravity \cite{CJS}. The renewed
interest in $D = 11$  supergravity has been appeared due to
unification of string theories in terms of the conjectured
$M$-theory \cite{Town,Witt}. Supergravity solutions with less than
maximal supersymmetry  may be of interest also in a context of
investigations  of  dual strongly coupled field theories \cite{KS}
and possible  modeling of multiple $M2$ and $M5$ configurations
\cite{BL1,BL2,Gust,ABJM,TY}.

Here we deal with non-marginal intersecting $M$-brane solutions
defined on product of Ricci-flat manifolds \cite{IM0,IM01}

 \beq{0.1}
 M_{0}  \times M_{1} \times \ldots \times M_{n}.
 \eeq
 These solutions are governed by several harmonic
functions defined on the manifold $M_0$ which is called as
transverse space  (see also \cite{IMC,IMtop,IM-sigma} for more
general solutions with composite branes on product of Ricci-flat
manifolds). Here we extend  an approach suggested earlier in
\cite{Iv-00} where  supersymmetric configurations with one brane
($M2$ and $M5$) on product of two Ricci-flat spaces were
considered.

We note that for flat  factor spaces $M_i = \R^{k_i} $, $i = 0,
\dots, n$, the supersymmetric  solutions with intersecting
$M$-branes were considered intensively in numerous publications,
see \cite{DS}-\cite{G} and references therein. The basic
$M2$-brane \cite{DS} and $M5$-brane \cite{Gu} solutions defined on
$\R^{k_0} \times \R^{k_1}$ (with $(k_0, k_1) = (8,3), (5,6)$,
respectively)  preserve     $1/2$ of supersymmetries (SUSY). The
classification of supersymmetric $M$-brane configurations on
product of flat factor spaces $M_i = \R^{k_i}$ was done in the
work of E. Bergshoeff {\it et al} \cite{BREJS}. The fractional
numbers of preserved SUSY are given by the relation \cite{BREJS}

 \beq{0.2}
 {\cal N} = 2^{-k}
 \eeq
 with $k = 1,2,3,4,5$.  (We note that in proving (\ref{0.2})
 the so-called ``$2^{-k}$-splitting'' theorem  from
 \cite{Iv-00} may be used.)

 The relation (\ref{0.2}) is no more valid if  composite
 $M$-brane configurations on product of Ricci-flat manifolds
 (\ref{0.1}) are taken into consideration. In this case ${\cal N}$
 depends upon certain numbers of  chiral parallel (i.e.
 covariantly constant)  spinors on $M_i$ and brane sign factors
 $c_s$. Here a composite Ansatz for the 4-form corresponding to $m$
 intersecting $M$-branes

 \beq{0.3}
          F  = \sum_{s=1}^{m} c_s {\cal F}_s,
 \eeq
 is assumed, where ${\cal F}_s$ is an elementary 4-form
corresponding to $s$-th $M$-brane and $c_s = \pm 1$  is the sign
(charge density) factor of $s$-th brane, $s = 1, ..., m$.

 For $M2$-brane solution on product of
 two Ricci-flat factor spaces $M_{0}  \times M_{1}$ ($n =1$, $m =1$)
 the number of unbroken SUSY is (at least) \cite{Iv-00}

 \beq{0.4} {\cal N} = n_0(c) n_1/32, \eeq
where  $n_0(c)$ is the number of chiral parallel spinors  on
8-dimensional transverse manifold $M_0$  with the chirality $c =
\pm 1$,   $n_1$ is the number of parallel spinors on 3-dimensional
world-volume $M_1$ and the chirality  number $c = \pm 1$ coincides
with the $M2$-brane sign. \fnm[2]\fnt[2]{Here and in what follows
the phrase ``... is the number of ... spinors'' means ``... is the
number of linear independent... spinors'' (or, more rigorously,
``... is the dimension of space of ... spinors'').}

 For $M_1 = \R^3$ relation (\ref{0.4}) was  used implicitly in
 \cite{DLPS} (for $M_0$ being the cones over certain 7-dimensional
 Einstein manifolds and  for $M_0 = \R_{+} \times M_{00}$, with certain Ricci-flat
   $M_{00}$) and in \cite{GGPT} (for hyper-K\"aler 8-dimensional manifold
  with holonomy group $Sp(2)$).

 For $M5$-brane solution on product of
 two Ricci-flat factor spaces $M_{0}  \times M_{1}$
 the number of unbroken SUSY is (at least) the following one
  \cite{Iv-00}

 \beq{0.5} {\cal N} = n_0 n_1(c)/32, \eeq
where  $n_0$ is the number of  parallel spinors on transverse
5-dimensional manifold $M_0$  and $n_1(c)$ is the number of chiral
parallel spinors on 6-dimensional world-volume manifold
 $M_1$. Here  the chirality $c = \pm 1$ is coinciding with the
 $M5$-brane sign.  Relation (\ref{0.5}) was  used  implicitly
  in  \cite{BP} (for  $M_0 = \R^5$),   \cite{K2}
 (for $M_1 = \R^2 \times K3$)  and \cite{FOF}
 (for $M_1$ with holonomy group $\R^4$ or $Sp(1) \ltimes \R^4$).

 Here we start  the  classification of partially supersymmetric
 $M$-brane  solutions defined on the products of Ricci-flat
  manifolds, i.e. we are aimed at the generalization of
 the  classification from \cite{BREJS}.

 The paper is organized as follows. In Section 2 we describe
 the basic notations  and definitions in arbitrary dimension $D$,
 e.g. splitting relations for spin  connection and spinorial covariant derivative on
 (warped) product  manifolds. Section 3 is devoted to SUSY ``Killing-like'' equations in
 11-dimensional supergravity and splitting procedure for the flux term
 in these equations.  In Section 4 the general class of
 non-marginal composite $M$-brane solutions  defined on product of
 Ricci-flat manifolds \cite{IM0,IM01} is considered. Here a key
 proposition concerning the solutions to SUSY  equations
 is presented.  This proposition (Proposition 1) is proved in Appendix.
 It gives a background for the calculation of the fractional numbers of
 preserved SUSY for  (non-marginal) composite M-brane solutions.
 In Section 5 the relations for these fractional numbers are derived
 for the following sets of brane:  $M2$, $M5$, $M2 \cap M2$,  $M2 \cap M5$,
 $M5 \cap M5$ and $M2 \cap M2 \cap M2$ and
  numerous examples for various factor-spaces $M_i$ are considered.

\section{\bf Basic notations}

Here we describe the basic notations in arbitrary dimension $D$.

\subsection{Product of manifolds}

Let  us consider the manifold

\beq{0.10} M = M_{0}  \times M_{1} \times \ldots \times M_{n},
\eeq with the metric \beq{0.11} g= e^{2{\gamma}(x)} \hat{g}^0  +
\sum_{i=1}^{n} e^{2\phi^i(x)} \hat{g}^i , \eeq

where $g^0  = g^0 _{\mu \nu}(x) dx^{\mu} \otimes dx^{\nu}$ is a
metric on the manifold $M_{0}$ and $g^i  = g^{i}_{m_{i}
n_{i}}(y_i) dy_i^{m_{i}} \otimes dy_i^{n_{i}}$ is a metric on  the
manifold $M_{i}$, $i = 1,\ldots, n$. Here and in what follows
$\hat{g}^{i}
 = p_{i}^{*} g^{i}$ is the pullback of the metric $g^{i}$  to the
manifold  $M$ by the canonical projection: $p_{i} : M \rightarrow
M_{i}$, $i = 0,\ldots, n$.

The functions $\gamma, \phi^{i} : M_0 \rightarrow \R$ are smooth.
We denote $d_{\nu} = {\rm dim} M_{\nu}$; $\nu = 0, \ldots, n$; $D
= \sum_{\nu = 0}^{n} d_{\nu}$. We put any  $M_{\nu}$, $\nu =
0,\ldots, n$, to be oriented and connected spin manifold.
 Then the volume
$d_i$-form
 \beq{0.14}
 \tau_i  \equiv \sqrt{|g^i(y_i)|}
 \ dy_i^{1} \wedge \ldots \wedge dy_i^{d_i},
 \eeq
is correctly defined for any $i=1,\ldots,n$.

Let $\Omega = \Omega(n)$  be a set of all non-empty subsets of $\{
1, \ldots,n \}$ ($|\Omega| = 2^n - 1$). For any $I = \{ i_1,
\ldots, i_k \} \in \Omega$, $i_1 < \ldots < i_k$, we denote
\bear{0.16}
 \tau(I) \equiv \hat{\tau}_{i_1}  \wedge \ldots \wedge \hat{\tau}_{i_k},  \\
   \label{0.19}
 d(I) \equiv  \sum_{i \in I} d_i.
\ear

Here and in what follows $\hat{\tau}_{i} = p_{i}^{*} \tau_{i}$ is
the pullback of the form $\tau_i$  to the manifold  $M$ by the
canonical projection: $p_{i} : M \rightarrow M_{i}$, $i =
1,\ldots, n$.

For $I \in \Omega$ we define an indicator of $i$ belonging to $I$

 \beq{0.20a}
 \delta_I^i \equiv \sum_{j \in
 I} \delta^i_j= \begin{array}{ll}
    1, & i  \in     I ,  \\
    0, & i  \notin  I .
\end{array}
\eeq

\subsection{Diagonalization of the metric}

For the metric $g  = g_{M N}(x) dx^{M} \otimes dx^{N}$ from
(\ref{0.11}), $M,N = 0, \ldots, D-1$, defined on the manifold
(\ref{0.10}), we define the diagonalizing $D$-bein $e^A = e^A_{\ \
M} dx^{M}$
 \beq{0.20}
 g_{MN} = \eta_{A B} e^A_{\ \ M} e^B_{\ \ N},
 \qquad \eta_{AB} = \eta^{AB} = \eta_A \delta_{AB},
 \eeq
$\eta_A = \pm 1$; $A,B = 0, \ldots, D-1$.

We choose the following frame vectors
 \beq{0.21}
 (e^A_{\ \ M}) = {\rm diag}(e^{\gamma} e^{(0) a}_{\ \ \ \ \mu},
  e^{ \phi^1} e^{(1) a_1}_{\ \ \ \ m_1}, \ldots,
  e^{\phi^n} e^{(n) a_n}_{\ \ \ \ m_n}),
 \eeq
where \beq{0.22}
 g^{0}_{\mu \nu} = \eta^{(0)}_{ab}
 e^{(0) a}_{\ \ \ \ \mu} e^{(0) b}_{\ \ \ \ \nu},
 \qquad
 g^{i}_{m_i n_i} = \eta^{(i)}_{a_i b_i}
 e^{(i) a_i}_{\ \ \ \ m_i} e^{(i) b_i}_{\ \ \ \ n_i},
\eeq
 $i = 1, \ldots, n$, and
\beq{0.23}
 (\eta_{AB}) = {\rm diag}(\eta^{(0)}_{ab},
  \eta^{(1)}_{a_1 b_1}, \ldots, \eta^{(n)}_{a_n b_n}).
\eeq For $(e^M_{\ \ A})  = (e^A_{\ \ M})^{-1}$ we get
 \beq{0.24}
 (e^M_{\ \ A}) = {\rm diag}(e^{- \gamma} e^{(0) \mu}_{\ \ \ \ a},
  e^{- \phi^1} e^{(1) m_1}_{\ \ \ \ a_1}, \ldots,
  e^{- \phi^n} e^{(n) m_n}_{\ \ \ \ a_n}),
\eeq where $(e^{(0) \mu}_{\ \ \ \ a}) = (e^{(0) a}_{\ \ \ \
\mu})^{-1}$,
 $(e^{(i) m_i}_{\ \ \ \ a_i}) = (e^{(i) a_i}_{\ \ \ \ n_i})^{-1}$,
 $i = 1, \ldots, n$.

{\bf Indices}. For indices we also use an alternative  numbering:
 $A = (a,a_1, \ldots, a_n)$, $B = (b,b_1, \ldots, b_n)$,
where $a,b = 1_0, \ldots, (d_0)_0$; $a_1,b_1 = 1_1, \ldots,
(d_1)_1$;
 ...; $a_n,b_n = 1_n, \ldots, (d_n)_n$; and
 $M = (\mu,m_1, \ldots, m_n)$,
 $N = (\nu,n_1, \ldots, n_n)$, where $\mu,\nu = 1_0, \ldots, (d_0)_0$;
 $m_1,n_1 = 1_1, \ldots, (d_1)_1$;
...; $m_n,n_n = 1_n, \ldots, (d_n)_n$.

\subsection{Gamma-matrices}

In what follows $\hat{\Gamma}_A$ are ``frame''
 $k \times k$  gamma-matrices satisfying
 \beq{0.25}
  \hat{\Gamma}_A \hat{\Gamma}_B + \hat{\Gamma}_B \hat{\Gamma}_A
  = 2 \eta_{AB} \1,
 \eeq
 $A,B = 0, \ldots, D-1$. Here $\1 = \1_k$ is unit $k \times k$ matrix and
 $k = 2^{[D/2]}$.

We also use ``world''  $\Gamma$-matrices
 \beq{0.26}
 \Gamma_M = e^A_{\ \ M} \hat{\Gamma}_A, \qquad
 \Gamma_M \Gamma_N + \Gamma_N \Gamma_M = 2 g_{MN} \1,
 \eeq
 $M,N = 0, \ldots, D-1$, and the matrices with upper indices:
 $\hat{\Gamma}^A=  \eta^{AB} \hat{\Gamma}_B$  and
 $\Gamma^M = g^{MN} \Gamma_N$. In what follows we will use the
 relation
 \beq{0.25h}
  \hat{\Gamma}^A \hat{\Gamma}^B + \hat{\Gamma}^B \hat{\Gamma}^A
  = 2 \eta^{AB} \1,
 \eeq
 where $\eta^{AB} = \eta_{AB}$.

 For any manifold $M_l$ with the metric $g^l$ we will also
 consider $k_l \times k_l$ $\Gamma$-matrices with $k_l = 2^{[d_l/2]}$
 obeying
   \beq{0.25l}
  \hat{\Gamma}^{a_l}_{(l)} \hat{\Gamma}^{b_l}_{(l)}  +
  \hat{\Gamma}^{b_l}_{(l)} \hat{\Gamma}^{a_l}_{(l)}
  = 2  \eta^{(l) a_l b_l} \1_{k_l}
  \eeq
  and
  \beq{0.25m}
  \hat{\Gamma}_{(l)} = \hat{\Gamma}^{1_l}_{(l)} \ldots
  \hat{\Gamma}^{(d_l)_l}_{(l)},
   \eeq
  $l = 0, \dots, n$.

\subsection{Spin connection}

Here we use the standard definition for the spin connection
 \beq{0.27}
 \omega^{A}_{\ \ BM} = \omega^{A}_{\ \ BM}(e,\eta)
 = e^A_{\ \ N}  \btd_M[g(e,\eta)] e^N_{\ \ B},
 \eeq
where the covariant derivative $\btd_M[g]$  corresponds to the
metric $g = g(e,\eta)$ from (\ref{0.20}). The spinorial covariant
derivative reads
 \beq{0.28}
 D_M = \p_M +  \frac{1}{4} \omega_{A B M} \hat{\Gamma}^A \hat{\Gamma}^B,
 \eeq
where $\omega_{A B M} = \eta_{AA'}\omega^{A'}_{\ \ BM}$,
$\omega_{A B M} =  - \omega_{B A M}$. For $D =4$ it was introduced
by V.A. Fock and D.D. Ivanenko in \cite{FockIv}.

The non-zero components of the spin connection (\ref{0.27}) in the
frame (\ref{0.21}) read
 \bear{0.29}
 \omega^{a}_{\ \ b \mu} =
 \omega^{a}_{\ \ b \mu}(e^{(0)},\eta^{(0)}) -
  e^{(0) \nu a} \gamma_{,\nu} e^{(0)}_{\ \ b \mu}
 + e^{(0) \nu}_{\ \ \ \ b} \gamma_{,\nu} e^{(0) a}_{\ \ \ \ \mu},
 \\  \label{0.30}
 \omega^{a}_{\ \ a_i m_j} = - \delta_{ij} e^{\phi^i - \gamma}
 (e^{(0) a}_{\ \ \ \ \nu} \btd^{\nu}[g^{(0)}] \phi^i ) e^{(i)}_{\ \ a_i
 m_i},
 \\  \label{0.31}
 \omega^{a_i}_{\ \ a m_j} = \delta_{ij} e^{\phi^i - \gamma}
 (e^{(0) \nu }_{\ \ \ \ a} \p_{\nu} \phi^i ) e^{(i) a_i}_{\ \ \ \ m_i},
 \\  \label{0.32}
 \omega^{a_i}_{\ \ b_j m_k} = \delta_{ij} \delta_{jk}
 \omega^{a_i}_{\ \ b_i m_i}(e^{(i)},\eta^{(i)}),
 \ear
 $i,j,k = 1, \ldots,n$, where $\omega^{a}_{\ \ b \mu}(e^{(0)},\eta^{(0)})$
and $\omega^{a_i}_{\ \ b_i m_i}(e^{(i)},\eta^{(i)})$ are
components  of the spin connections corresponding to the metrics
from
 (\ref{0.22}).

Let
 \beq{0.33}
 A_{M} \equiv \omega_{A B M} \hat{\Gamma}^A \hat{\Gamma}^B.
 \eeq
For  $A_M = A_M(e,\eta,\hat{\Gamma}^C)$ in the frame (\ref{0.21})
we get
 \bear{0.34}
 A_{\mu} = \omega^{(0)}_{a b \mu}  \hat{\Gamma}^a \hat{\Gamma}^b +
 (\Gamma_{\mu} \Gamma^{\nu} - \Gamma^{\nu} \Gamma_{\mu} ) \gamma_{,
 \nu},
 \\  \label{0.35}
 A_{m_i} = \omega^{(i)}_{a_i b_i m_i}  \hat{\Gamma}^{a_i}
 \hat{\Gamma}^{b_i} + 2 \Gamma_{m_i} \Gamma^{\nu} \phi^{i}_{, \nu},
 \ear
where $\omega^{(0)}_{a b \mu} = \omega_{a b
\mu}(e^{(0)},\eta^{(0)})$ and $\omega^{(i)}_{a_i b_i m_i} =
\omega_{a_i b_i m_i}(e^{(i)},\eta^{(i)})$,
 $i =1, \ldots, n$.

 Relations (\ref{0.34}) and (\ref{0.35}) imply the
 following decomposition of the covariant derivative (\ref{0.28})

  \bear{0.36}
 D_{\mu} = \bar{D}_{\mu}^{(0)}  + \frac{1}{4}
 (\Gamma_{\mu} \Gamma^{\nu} - \Gamma^{\nu} \Gamma_{\mu} ) \gamma_{,
 \nu},
 \\  \label{0.37}
  D_{m_i} = \bar{D}_{m_i}^{(i)} +  \frac{1}{2}\Gamma_{m_i} \Gamma^{\nu} \phi^{i}_{, \nu},
  \qquad i > 0,
 \ear
  where
   \beq{0.38}
   \bar{D}_{m_l}^{(l)} = \p_{m_l} + \frac{1}{4}
      \omega^{(l)}_{a_l b_l m_l}  \hat{\Gamma}^{a_l}
      \hat{\Gamma}^{b_l},
   \eeq
    $l = 0, ..., n$  ($\mu = m_0$).

   In what follows operators (\ref{0.38}) will generate
   the covariant spinorial derivatives corresponding the manifolds  $M_l$
    \beq{0.38a}
    D^{(l)}_{m_l} = \p_{m_l} +  \frac{1}{4} \omega^{(l)}_{a_l b_l m_l}
                    \hat{\Gamma}^{a_l}_{(l)} \hat{\Gamma}^{b_l}_{(l)},
    \eeq
    $l = 0, ..., n$.

 \section{SUSY equations}

We consider the $D =11$ supergravity with the action in the
bosonic sector  \cite{CJS}
 \beq{1.1}
  S= \int d^{11}z \sqrt{|g|} \biggl\{R[g] - \frac{1}{2 (4!)} F^2 \biggr\}
     - \frac{1}{6} \int A \wedge F \wedge F,
 \eeq
where  $F = d A$ is $4$-form. Here we consider pure bosonic
configurations in $D =11$ supergravity (with zero fermionic
fields) that are solutions to the equations of motion
corresponding to the action (\ref{1.1}).

The number of supersymmetries (SUSY) corresponding to the bosonic
background $(e^{A}_M, A_{M_1 M_2 M_3})$ is defined by the
dimension of the space of solutions to (a set of) linear
first-order differential equations (SUSY eqs.)
 \beq{1.2}
 (D_M  + B_M ) \varepsilon = 0,
 \eeq
where  $D_M$ is covariant spinorial derivative from (\ref{0.28}),
$\varepsilon = \varepsilon (z)$  is
 $32$-component    spinor field (see Remarks 1 and 2 below)
 and
 \beq{1.3}
 B_M  = \frac{1}{288}
 (\Gamma_M \Gamma^N \Gamma^P \Gamma^Q \Gamma^R -
 12  \delta_M^N  \Gamma^P \Gamma^Q \Gamma^R) F_{NPQR}.
 \eeq
Here  $F =  dA = \frac{1}{4!} F_{NPQR} dz^{N} \wedge dz^{P} \wedge
dz^{Q} \wedge dz^{R}$, and $\Gamma_M$ are world $\Gamma$-matrices.

The number of unbroken SUSY is
 \beq{1.4}
  {\cal N} = N/32,
 \eeq
 where $N$ is the dimension of linear space of solutions to
  differential equations (\ref{1.2}).

{\bf Remark 1.} In this paper we put for simplicity $\varepsilon
(z) \in \C^{32}$. The imposing of Majorana condition
$\bar{\varepsilon}
    = B \varepsilon$, where $\bar{(.)}$ denotes the complex conjugation and
      $B$ is non-degenerate matrix obeying
    $\bar{\hat{\Gamma}}_A = \mp B \bar{\hat{\Gamma}}_A B^{-1}$,
    will give the same number $N$ for the dimension of real linear
    space of parallel Majorana spinors obeying (\ref{1.2}).

{\bf Remark 2.} A possible  consistent approach to superanalysis
implies the use of infinite-dimensional super-commutative Banach
algebras, e.g. Grassmann-Banach ones \cite{Rog,VV,Khr1,I1} (see
also \cite{DeWitt}). One may put $\varepsilon (z) \in
(\G_{\1})^{32}$, where $\G_{\1}$ is an odd part of an
infinite-dimensional Grassmann-Banach algebra
 $\G = \G_0 \oplus \G_1$ \cite{Rog,I1,Iv}. In
this case the complex-valued solution $\varepsilon(z)$ should be
replaced by $\varepsilon(z) g_1$, where $g_1$ is arbitrary element
of $\G_1$.

Here we consider the  decomposition  of matrix-valued  field $B_M$
on the product manifold (\ref{0.10}) in the frame (\ref{0.21}) for
electric and magnetic branes.

{\bf $M2$-brane.} Let the $4$-form be

  \beq{2.1}
  F = d \Phi \wedge \tau(I)
  \eeq
  where $\Phi = \Phi(x)$,
 $I = \{ i_1, \ldots, i_k \}$, $i_1 < \ldots < i_k$, $d(I) =3$. The
 calculations give \cite{Iv-00}

 \bear{2.2}
 B_{m_l} = \frac{1}{12} S(I) \exp(- \sum_{i \in
  I} d_i \phi^i ) [(1 - 3 \delta_I^l) \Gamma_{m_l} \Gamma^{\nu}
  \Phi_{, \nu} - 3 \delta_0^l \Phi_{, m_l} ] \hat{\Gamma}(I),
 \ear
 $l = 0, \ldots, n$, where  $m_0 = \mu$,
  \beq{2.2s}
   S(I) = {\rm sign}(\prod_{i
          \in I} {\rm det} (e^{(i) m_i}_{\ \ \ \ a_i}))
   \eeq
 and

  \beq{2.2a}
  \hat{\Gamma}(I) = \hat{\Gamma}^{A_1} \hat{\Gamma}^{A_2}
  \hat{\Gamma}^{A_3}
  \eeq
 with $(A_1, A_2, A_3) =
       (1_{i_1}, \ldots, (d_{i_1})_{i_1}, \ldots, 1_{i_k}, \ldots,
       (d_{i_k})_{i_k})$.

 {\bf $M5$-brane.} Let

  \beq{2.3}
   F =  (*_0 d \Phi) \wedge    \tau(\bar{I}),
  \eeq
  where $*_0$ is the Hodge operator on
  $(M_0,g^0)$ and $\bar{I} = \{1, \dots, n \} \setminus I = \{ j_1,
  \ldots, j_l \}$, $j_1 < \ldots < j_l$. It follows from (\ref{2.3})
 that $d_0 + d(\bar{I}) = 5$ and $d(I) = 6$. We get \cite{Iv-00}

 \bear{2.4}
 B_{m_l} = \frac{1}{24} S(\{ 0 \}) S(\bar{I})\exp[- (d_0
  -2) \gamma -  \sum_{i \in \bar{I} } d_i \phi^i ] \times
  \qquad \\ \nn
 \times [2 \Gamma_{m_l} \Gamma^{\nu} \Phi_{, \nu} - 3 \delta_0^l
  (\Gamma_{m_l} \Gamma^{\nu} - \Gamma^{\nu} \Gamma_{m_l}) \Phi_{,
  \nu} + 6 \delta_{\bar{I}}^l \Gamma^{\nu} \Gamma_{m_l}) \Phi_{,
 \nu}] \hat{\Gamma}(\{ 0 \}) \hat{\Gamma}(\bar{I}),
 \ear
  $l = 0, \ldots, n$, where

  \beq{2.4s}
  S(\{0 \}) = {\rm sign}({\rm det} (e^{(0) \nu}_{\ \
                \ \ a})),
  \eeq
  and
 \bear{2.4a}
 \hat{\Gamma}(\{0 \}) = \hat{\Gamma}^{1_0} \ldots
 \hat{\Gamma}^{(d_0)_0}, \\
 \label{2.4b}
 \hat{\Gamma}(\bar{I}) =
 \hat{\Gamma}^{B_1} \ldots \hat{\Gamma}^{B_k}
 \ear
 with
 $(B_1, \ldots, B_k) = (1_{j_1}, \ldots, (d_{j_1})_{j_1},
 \ldots, 1_{j_l}, \ldots, (d_{j_l})_{j_l})$ and $d_0 + k = 5$.

 \section{Composite M-brane solutions}

 We consider a classical solution corresponding
 to the action (\ref{1.1}) \cite{IM0,IM01,IMC}.
 The metric is  defined on the  product manifold  (\ref{0.10})
 and has the following form

  \bear{2.5}
   g= e^{2{\gamma}(x)} \hat{g}^0  +
   \sum_{i=1}^{n} e^{2\phi^i(x)} \hat{g}^i ,
   \\ \label{2.5a}
   e^{2{\gamma}} = (\prod_{s \in S_e} H_s )^{1/3}
                     (\prod_{s \in S_m} H_s )^{2/3},
    \\ \label{2.5b}
   e^{2\phi^i} =  e^{2{\gamma}} \prod_{s \in S} H_s^{ - \delta^i_{I_s}},
  \ear
  $i = 1, \dots, n$.

 Here $g^0  = g^0 _{\mu \nu}(x) dx^{\mu} \otimes dx^{\nu}$ is a
 Ricci-flat metric on the manifold $M_{0}$ and $g^i  = g^{i}_{m_{i}
 n_{i}}(y_i) dy_i^{m_{i}} \otimes dy_i^{n_{i}}$ is a Ricci-flat metric on
 $M_{i}$, $i = 1,\ldots, n$.

  The 4-form reads

 \bear{2.6}
  F = \sum_{s \in S_e} c_s d H_s^{-1} \wedge \tau (I_s)
   + \sum_{s \in S_m} c_s
           (*_0 d H_s) \wedge  \tau (\bar I_s),
 \ear
 where $c^2_s =  1$;
 $*_0$ is the Hodge operator on $(M_0,g^0)$, $H_s$ are harmonic functions
 on $(M_0,g^0)$ and

  \beq{2.7}
  \bar I_s = \{1, ..., n \} \setminus I_s
  \eeq
  is dual set.

 Here  the set of indices $S_e \subset \Omega(n)$
 describes electric branes with worldvolume dimensions $d(I_s) =
 3$, $I_s \in S_e$ and $S_m \subset \Omega(n)$
 describes magnetic branes with worldvolume dimensions $d(I_s) =
 6$, $I_s \in S_m$. The intersections rules are standard ones
       \beq{2.8}
        d(I_s \cap I_{s'}) = \frac{1}{9} d(I_s) d(I_{s'}),
       \eeq
 $s \neq s'$, $s, s' \in S$, where $S = S_e \cup S_m$, or,
 explicitly,
   \beq{2.8a}
        d(I_s \cap I_{s'}) = 1, 2, 4,
       \eeq
 for $M2 \cap M2$, $M2 \cap M5$ and $M5 \cap M5$ cases, respectively.

 Only one manifold $(M_{i_0}, g^{i_0})$ ($i_0 > 0$)  has a pseudo-Euclidean
 signature while all others $(M_{i}, g^{i})$ ($i \neq i_0$)
 should be of Euclidean signature. The index $i_0$ is
 contained by all brane sets: $i_0 \in I_s$, $s \in S$.

It should be noted that here the Chern-Simons term in the action
(\ref{1.1}) does not give the contribution into equations of
motion due to
        \beq{2.8cs}
        F \wedge F =0.
        \eeq

 Using the  decomposition  of matrix-valued  field $B_M$ on the
 product manifold (\ref{0.10}) in the frame (\ref{0.21}) for
 electric and magnetic branes from (\ref{2.2}) and (\ref{2.4}) we
 obtain
     \beq{2.9}
      B_M = \sum_{s \in S} B_M^s,
     \eeq
  where
  \bear{2.10}
  B_{m_l}^s = \frac{c_s}{12} H_s
  [ (1 - 3 \delta_{I_s}^l) \Gamma_{m_l} \Gamma^{\nu} \p_{\nu}
  H_s^{-1}    - 3 \delta_0^l \p_{m_l} H_s^{-1} ] \hat{\Gamma}_{[s]},
  \ear
   for $s \in S_e$, and

 \bear{2.11}
  B_{m_l}^s = \frac{c_s}{24} H_s^{-1}
  [ 2 \Gamma_{m_l} \Gamma^{\nu} \p_{\nu} H_s - 3 \delta_0^l
  (\Gamma_{m_l} \Gamma^{\nu} - \Gamma^{\nu} \Gamma_{m_l}) \p_{\nu} H_s
  + 6 \delta_{\bar{I}_s}^l \Gamma^{\nu} \Gamma_{m_l} \p_{\nu} H_s ] \hat{\Gamma}_{[s]},
 \ear
  for $s \in S_m$;  $l = 0, \ldots, n$.

 Here we denote
    \bear{2.12}
        c_{[s]} = c_s  S(I_s), \qquad s \in S_e; \\ \nn
        c_{[s]} = c_s  S(\{ 0 \}) S(\bar{I_s}), \qquad  s \in S_m .
     \ear
     and
      \bear{2.13e}
        \hat{\Gamma}_{[s]} =  \hat{\Gamma}(I_s), \qquad s \in S_e; \\
         \label{2.13m}
          \hat{\Gamma}_{[s]} = \hat{\Gamma}(\{ 0 \}) \hat{\Gamma}(\bar{I_s}), \qquad  s \in S_m.
     \ear

 In derivation of (\ref{2.9})-(\ref{2.11}) the following formulae
 are used
     \bear{2.15a}
    - \sum_{i \in I_s} d_i \phi^i =  \ln H_s, \qquad s \in S_e;
       \\ \label{2.15b}
    - \gamma (d_0 -2)   -
       \sum_{i \in \bar{I}_s} d_i \phi^i = -  \ln H_s, \qquad  s \in S_m.
     \ear
 The proof of these  relations  is given in Appendix A.
 It is based on intersection rules  (\ref{2.8}).

    The definitions (\ref{2.13e}) and (\ref{2.13m}) imply
      \beq{2.14e}
          \hat{\Gamma}_{[s]} =  \hat{\Gamma}^{A_1}  \hat{\Gamma}^{A_2} \hat{\Gamma}^{A_3},
         \qquad {\rm for} \ s \in S_e,
      \eeq
      where $(\hat{\Gamma}^{A_{i_0}})^2 = - \1$ for some $i_0 \in
      \{1,2,3\}$, $(\hat{\Gamma}^{A_{i}})^2 =  \1$ for $i \neq i_0$,
       and
      \beq{2.14m}
          \hat{\Gamma}_{[s]} = \hat{\Gamma}^{B_1}  \hat{\Gamma}^{B_2} \hat{\Gamma}^{B_3}
         \hat{\Gamma}^{B_4}  \hat{\Gamma}^{B_5}, \qquad  {\rm for} \ s \in S_m,
      \eeq
      where  $(\hat{\Gamma}^{B_i})^2 =  \1$ for all $i$.

      It  follows  from the relations (\ref{2.13e}) and (\ref{2.13m}),
      that
       \beq{2.14p}
       ( \hat{\Gamma}_{[s]})^2 = \1,
       \eeq
       $\1 = \1_{32}$. The matrices $\hat{\Gamma}_{[s]}$ commute
       with each other due to intersection rules (\ref{2.8a}).

      In what follows the following proposition plays a key role.

    {\bf Proposition 1.} { \it Let
      \beq{2.16}
       \varepsilon = (\prod_{s \in S_e} H_s)^{-1/6} (\prod_{s \in S_m} H_s)^{- 1/12} \eta,
       \eeq
     where
     \beq{2.16a}
      \bar{D}_{m_l}^{(l)} \eta = 0, \qquad l = 0, \dots, n,
      \eeq
      and
     \beq{2.16b}
       \hat{\Gamma}_{[s]} \eta = c_{[s]} \eta
      \eeq
      for all $s \in S$. Then the relation (\ref{1.2})

       $$(D_M  + B_M ) \varepsilon = 0$$

       is satisfied identically. }

       The proof of this proposition is given in Appendix B.
       For flat   factor-spaces $M_i = \R^{k_i}$
       (and suitably chosen beins  $(e^{(i) m_i}_{\ \ \ \ a_i})$),
         $i = 0, \ldots, n$,  relations (\ref{2.16a}) are satisied identically
         for constant $\eta$  and hence we should  deal only with the set of equations
         (\ref{2.16b}) \cite{BREJS}.

       When all $H_s = 1$ and $F = 0$ (i.e. all branes are removed)  we
       get a  more simple proposition.

          {\bf Proposition 2.} { \it  $\eta$ is parallel spinor (i.e. $D_M \eta = 0$)
          on the product manifold (\ref{0.1})  if and only if
              \beq{2.16c}
          \bar{D}_{m_l}^{(l)} \eta = 0, \qquad l = 0, \dots, n.         \eeq
            }

          Proposition 2 may be used for constructing chiral parallel spinors on product manifolds.
          An example decribing  chiral parallel spinors on product of two 4-dimensional manifolds
          (of Euclidean signatures) is given Appendix C.

 \section{Supersymmetric composite $M$-brane solutions}

Here we consider certain examples of supersymmetric solutions. We
put for simplicity that
   \beq{2.17}
  {\rm det} (e^{(l) m_l}_{\ \ \ \ a_l})) > 0,
   \eeq
  $l = 0, \dots, n$, and hence
  \beq{2.18}
  c_{[s]} = c_s
  \eeq
  for all $s \in  S$
  (see (\ref{2.2s}), (\ref{2.4s}) and (\ref{2.12})).

 \subsection{$M2$-brane}

Let us we consider (in detail) the electric  $2$-brane solution
defined on the manifold

 \beq{3.2a} M_{0}  \times M_{1}. \eeq

The solution reads

 \bear{3.2}
  g= H^{1/3} \{ \hat{g}^0  + H^{-1} \hat{g}^1   \},
  \\ \label{3.3}
  F = c d H^{-1} \wedge \hat{\tau}_1,
 \ear
 where  $c^2 = 1$, $H = H(x)$ is a harmonic function on
 $(M_0,g^0)$,  $d_1 = 3$, $d_0 = 8$ and the metrics $g^i$, $i =
  0,1$, are Ricci-flat; $g^0$ has the Euclidean signature and $g^1$ has
 the signature $(-,+,+)$.

We consider $\Gamma$-matrices

  \beq{3.3n}
   (\hat{\Gamma}^A) = (\hat{\Gamma}^{a_0}_{(0)} \otimes
  \1_2, \hat{\Gamma}_{(0)} \otimes \hat{\Gamma}^{a_1}_{(1)}),
  \eeq

 where $16 \times 16$ $\Gamma$-matrices $\hat{\Gamma}^{a_0}_{(0)}$, $a_0 =
 1_0,  \ldots, 8_0$, correspond to $M_0$:

 \beq{3.4}
 \hat{\Gamma}^{a_0}_{(0)} \hat{\Gamma}^{b_0}_{(0)} +
 \hat{\Gamma}^{b_0}_{(0)} \hat{\Gamma}^{a_0}_{(0)}  = 2 \delta_{a_0
 b_0} \1_{16},
 \eeq
  $2 \times 2$ $\Gamma$-matrices
 $\hat{\Gamma}^{a_1}_{(1)}$,   $a_1 = 1_1, 2_1, 3_1$,
 correspond to $M_1$:
  \beq{3.5}
 \hat{\Gamma}^{a_1}_{(1)} \hat{\Gamma}^{b_1}_{(1)} +
 \hat{\Gamma}^{b_1}_{(1)} \hat{\Gamma}^{a_1}_{(1)}  = 2
 \eta_{a_1  b_1} \1_{2}
 \eeq
 with  $(\eta_{a_1  b_1}) = {\rm diag}(-1,+1,+1)$
  and $\hat{\Gamma}_{(0)}
  =   \hat{\Gamma}^{1_0}_{(0)} \ldots \hat{\Gamma}^{8_0}_{(0)}$.

 The  $\Gamma$-matrices (\ref{3.3n}) obey the relations (\ref{0.25h})
 due to (\ref{3.4}), (\ref{3.5}) and the identity
  \beq{3.5a}
 (\hat{\Gamma}_{(0)})^2 =  \1_{16}.
  \eeq

 It follows from  (\ref{0.38}),  (\ref{3.3n}) and (\ref{3.5a}) that
    \bear{3.6a}
     \bar{D}_{m_0}^{(0)} = \p_{m_0} + \frac{1}{4}
     \omega^{(0)}_{a_0 b_0 m_0}  \hat{\Gamma}^{a_0}
     \hat{\Gamma}^{b_0} \otimes \1_2, \\ \label{3.6b}
     \bar{D}_{m_1}^{(1)} = \p_{m_1} + \frac{1}{4}
     \omega^{(1)}_{a_1 b_1 m_1} \1_{16} \otimes \hat{\Gamma}^{a_1}
     \hat{\Gamma}^{b_1}.
   \ear

   Let
   \beq{3.6c}
   \eta =  \eta_0(x) \otimes \eta_1(y_1),
   \eeq
    where $\eta_0  = \eta_0(x)$ is $16$-component spinor on $M_0$,
  and   $\eta_1 =  \eta_1(y_1)$ is $2$-component spinor on
   $M_1$. Then we get from (\ref{3.6a}) and (\ref{3.6b})
    \beq{3.7}
     \bar{D}_{m_0}^{(0)} \eta  =  (D_{m_0}^{(0)} \eta_0) \otimes \eta_1,
     \qquad
     \bar{D}_{m_1}^{(1)} \eta =     \eta_0   \otimes (D_{m_1}^{(1)}
     \eta_1),
     \eeq
  where $D^{(0)}_{m_0} = \p_{m_0} +  \frac{1}{4} \omega^{(0)}_{a_0 b_0 m_0}
  \hat{\Gamma}^{a_0}_{(0)} \hat{\Gamma}^{b_0}_{(0)}$ and $D^{(1)}_{m_1} =
  \p_{m_1} +  \frac{1}{4} \omega^{(1)}_{a_1 b_1 m_1}
   \hat{\Gamma}^{a_1}_{(1)} \hat{\Gamma}^{b_1}_{(1)}$
   are covariant (spinorial) derivatives (\ref{0.38a}) corresponding to the manifolds
   $M_0$ and $M_1$, respectively.

  The  operator (\ref{2.14e}) corresponding to $M2$-brane reads

   \beq{3.8a}
     \hat{\Gamma}_{[s]} =  \hat{\Gamma}^{1_1}  \hat{\Gamma}^{2_1} \hat{\Gamma}^{3_1} =
                   \hat{\Gamma}_{(0)} \otimes \hat{\Gamma}_{(1)},
     \eeq
  where  $\hat{\Gamma}_{(1)} = \hat{\Gamma}^{1_1}_{(1)}
  \hat{\Gamma}^{2_1}_{(1)} \hat{\Gamma}^{3_1}_{(1)}$.

 Choosing (real) matrices
  \beq{3.8b}
  \hat{\Gamma}^{1_1}_{(1)} = i \sigma_2, \quad
  \hat{\Gamma}^{2_1}_{(1)} =  \sigma_1, \quad
  \hat{\Gamma}^{3_1}_{(1)} =  \sigma_3 =
  \hat{\Gamma}^{1_1}_{(1)} \hat{\Gamma}^{2_1}_{(1)},
   \eeq
 where $\sigma_i$ are the standard Pauli matrices, we get
 \beq{3.8c}
 \hat{\Gamma}_{(1)} =  \1_2.
 \eeq

 Due to (\ref{3.6c}), (\ref{3.8a}) and (\ref{3.8c}) the
 chirality restriction (\ref{2.16b}) in  Proposition 1
  is satisfied if
  \beq{3.8d}
    \hat{\Gamma}_{(0)} \eta_0  = c \eta_0,
     \eeq
see (\ref{2.18}).

 Using the Proposition 1 we are led to the following
 solution to SUSY equations (\ref{1.2})
 corresponding to the field configuration from
(\ref{3.2}), (\ref{3.3})
 \beq{3.8e}
 \varepsilon  = H^{-1/6} \eta_0(x) \otimes \eta_1(y).
 \eeq

 Here  $\eta_0(x)$ is a $16$-component parallel  (Killing) chiral spinor
 (field) on $M_0$ ($D^{(0)}_{m_0} \eta_0 = 0$ )
 obeying (\ref{3.8d}) and $\eta_1(y)$ is a
 $2$-component parallel spinor  on $M_1$
  ($D^{(1)}_{m_1} \eta_1 = 0$).

 Hence the number of unbroken SUSY is at least \cite{Iv-00}

 \beq{3.10} {\cal N} = n_0(c) n_1/32, \eeq

where  $n_0(c)$ is the number of chiral parallel (Killing) spinors
on $M_0$ satisfying (\ref{3.8d}),  and $n_1$ is the number of
parallel spinors on $M_1$.

Since the $3$-dimensional space $(M_1,g^1)$ is considered to be
Ricci-flat,  it is flat, i.e. the Riemann tensor corresponding to
 $g^1$ is zero.

 {\bf Case $n_1 = 2$ }. Let us consider flat pseudo-Euclidean space
 $M_1 = \R^3$, $g^1 = - d y^1_1
 \otimes d y^1_1 + d y^2_1 \otimes d y^2_1 + d y^3_1 \otimes d
  y^3_1 $. In this case $n_1 = 2$ and

 \beq{3.11} {\cal N} = n_0(c) /16. \eeq

For flat $M_0 = \R^8$ we get $n_0(c) =  8$ and hence ${\cal N} =
1/2$ in agreement with \cite{DS}.

 {\bf Example:  $M_0$ with holomomy groups  $Spin(7)$, $SU(4)$, $Sp(2)$. }
According to M. Wang's classification \cite{Wang}  an irreducible,
simply-connected, Riemannian 8-dimensional manifold $M_0$
admitting parallel spinors must have precisely one of the
following holonomy groups: $Spin(7)$, $SU(4)$ or $Sp(2)$ . For
suitably chosen  orientation on $M_0$ the numbers of chiral
parallel spinors are the following ones $(n_0(+1), n_0(-1)) = (k,
0)$, where $k = 1, 2, 3$ for $H  = Spin(7), SU(4),  Sp(2)$,
respectively. Hence  ${\cal N} = k /16$ for $c = + 1$ and ${\cal
N}  = 0$ for $c = -1$. The  hyper-K\"aler case with $H = Sp(2)$
 was studied in \cite{GGPT}.

 {\bf Example: $M_0$ with holonomy groups $SU(2)$ and $SU(2) \times SU(2)$. }
Let us consider $K3 = CY_2$ which is a 4-dimensional  Ricci-flat
K\"{a}hler manifold with a holonomy group $SU(2) = Sp(1)$ and
self-dual (or anti-self-dual) curvature tensor. $K3$ has two
Killing spinors of the same chirality, say, $+1$.   Let $M_0 =
\R^4 \times K3$, then we get $n_0 (+1) = n_0 (-1) = 4$ and hence
 ${\cal N} =  1/4$ for any $c = \pm 1$. (See Appendix C.)
  For  $M_0 = K3 \times K3$,  we obtain $n_0 (+1) = 4$,  $n_0 (-1) = 0$
 and hence ${\cal N} =  1/4$ for  $c = 1$ and ${\cal N} = 0$  for  $c = -1$ .

{\bf Example: $M_0$ with holonomy group $SU(3)$. }
 Let $M_0 = \R^2 \times CY_3$ be 6-dimensional Calabi-Yau manifold (3-fold) of
 holonomy $SU(3)$. Since $CY_3$ has two parallel spinors of
 opposite chiralities (and  the same for $\R^2$)  we  get $n_0 (+1) = n_0 (-1) = 2$
  and hence ${\cal N} =  1/8$ for any $c = \pm 1$.

\subsection{$M5$-brane}

Now we consider the magnetic  $5$-brane solution defined on the
manifold  (\ref{3.2a}) with $d_0 =5$ and $d_1 = 6$:

 \bear{3.4m} g= H^{2/3} \{ \hat{g}^0  + H^{-1} \hat{g}^1 \},
 \\ \label{3.5m}
 F = c (*_0 d H),
 \ear
 where $c^2 = 1$,
 $*_0$ is the Hodge operator on $(M_0,g^0)$,
 $H = H(x)$ is a harmonic function on
$(M_0,g^0)$,  and metrics $g^i$, $i = 0,1$, are Ricci-flat, $g^0$
has a Euclidean signature and $g^1$ has the signature
$(-,+,+,+,+,+)$.

Let us consider $\Gamma$-matrices

 \beq{3.3nm}
  (\hat{\Gamma}^A) = (\hat{\Gamma}^{a_0}_{(0)} \otimes
  \hat{\Gamma}_{(1)}, \1_4 \otimes \hat{\Gamma}^{a_1}_{(1)}),
 \eeq
where
 $\hat{\Gamma}^{a_0}_{(0)}$, $a_0 = 1_0, \ldots, 5_0$,
 are $4 \times 4$ $\Gamma$-matrices
corresponding to $M_0$ and $\hat{\Gamma}^{a_1}_{(1)}$, $a_1 = 1_1,
\ldots, 6_1$, are $8 \times 8$  $\Gamma$-matrices corresponding to
$M_1$. Here  $\hat{\Gamma}_{(1)} =
  \hat{\Gamma}^{1_1}_{(1)} \ldots \hat{\Gamma}^{6_1}_{(1)}$
  and $(\hat{\Gamma}_{(1)})^2 = \1_8$.

 We put $\eta =  \eta_0(x) \otimes \eta_1(y_1)$,
   where $\eta_0  = \eta_0(x)$ is $4$-component spinor on $M_0$,
  and   $\eta_1 =  \eta_1(y_1)$ is $8$-component spinor on
   $M_1$. Then relations (\ref{3.7}) are  valid   in magnetic case too.

The operator (\ref{2.14m}) corresponding to $M5$-brane reads
   \beq{3.8am}
     \hat{\Gamma}_{[s]} =  \hat{\Gamma}^{1_0}  \hat{\Gamma}^{2_0} \hat{\Gamma}^{3_0}
                  \hat{\Gamma}^{4_0} \hat{\Gamma}^{5_0}
      =    \hat{\Gamma}_{(0)} \otimes \hat{\Gamma}_{(1)},
     \eeq
  where  $\hat{\Gamma}_{(0)} = \hat{\Gamma}^{1_0}_{(0)} \dots
   \hat{\Gamma}^{5_0}_{(0)} = \1_4$, if we put
   $\hat{\Gamma}^{5_0}_{(0)} = \hat{\Gamma}^{1_0}_{(0)} \dots
   \hat{\Gamma}^{4_0}_{(0)}$. Then the
 chirality restriction (\ref{2.16b})
 is satisfied if
   \beq{3.8dm}
    \hat{\Gamma}_{(1)} \eta_1  = c \eta_1.
     \eeq

Using the Proposition 1 we get the following
 solution to  equations (\ref{1.2})
 corresponding to the field configuration from
(\ref{3.4m}) and  (\ref{3.5m})
 \beq{3.8em}
 \varepsilon  = H^{-1/12} \eta_0(x) \otimes \eta_1(y).
  \eeq
 Here  $\eta_0(x)$ is a $4$-component parallel (Killing) spinor
 (field) on $M_0$ ($D^{(0)}_{m_0} \eta_0 = 0$) and $\eta_1(y)$ is a
 $8$-component chiral parallel  spinor  on $M_1$
  ($D^{(1)}_{m_1} \eta_1 = 0$) obeying  (\ref{3.8dm}).

 Hence the number of unbroken SUSY is at least \cite{Iv-00}

 \beq{3.10m} {\cal N} = n_0 n_1(c)/32, \eeq

where  $n_0$ is the number of  parallel spinors on $M_0$ , and
$n_1(c)$ is the number of chiral parallel spinors on $M_1$
satisfying (\ref{3.8dm}).

 For flat factor-spaces with $M_0 = \R^5$ (and Euclidean metric) and
 $M_1 = \R^6$ (and Minkowskian metric of $\R^{1,5}$) we get
 $n_0 = 4$, $n_1(c) =4$ and hence ${\cal N} = 1/2$ in agreement
 with \cite{Gu}.

{\bf Example: generalized Kaya solution.} Let $M_1 = \R^2 \times
 K3$. In this case  we obtain from (\ref{3.10m})
 $${\cal N} = n_0/16$$   since $n_1(c) = 2$.
 For flat $M_0 = \R^5$ ($n_0 =
 4$) we get   ${\cal N} = 1/4$  \cite{K2}. For $M_0 = \R \times K3$
 we have $n_0 = 2$ and consequently ${\cal N} = 1/8$.

{\bf Example: generalized Figueroa-O'Farrill solutions.} Let
 $(M_1,g^1)$ be a 6-dimensional Ricci-flat $pp$-wave solution from
 \cite{FOF} (page. 5) of holonomy   $Sp(1) \ltimes \R^4$ or  $\R^4$
 for $k = 1, 2$, respectively.  Then the numbers of chiral parallel spinors are
 $(n_1(+1), n_1(-1)) = (k,k)$, $k = 1,2$. Due to
 (\ref{3.10m}) we have

  $${\cal N} = n_0 k/32.$$

 For flat $M_0 = \R^5$ we have
   ${\cal N} = 1/8, 1/4$ for $k = 1, 2$, respectively, in agreement with
   \cite{FOF}. For $M_0 = \R \times K3$ we reduce these numbers to ${\cal N} = 1/16, 1/8$,
   for $k = 1, 2$, respectively.

\subsection{$M2 \cap M5$-branes}

Let us consider a solution  with intersecting $M2$- and
$M5$-branes defined on the manifold

 \bear{4.1}
 M_{0}  \times M_{1}  \times M_{2} \times M_{3},
  \ear
 where $d_0 = 4$, $d_1 = 1$, $d_2 = 4$ and $d_3 = 2$.

 The solution reads

 \bear{4.2}
 g= H_1^{1/3} H_2^{2/3}
 \{ \hat{g}^0  + H_1^{-1} \hat{g}^1 +  H_2^{-1} \hat{g}^2
 + H_1^{-1} H_2^{-1} \hat{g}^3 \},
 \\ \label{4.3}
 F = c_1 d H^{-1}_1 \wedge  \hat{\tau}_1 \wedge \hat{\tau}_3 +
     c_2 (*_0 d H_2) \wedge \hat{\tau}_1,
 \ear

where $c^2_1 = c^2_2 = 1$; $H_1, H_2$ are harmonic functions on
$(M_0,g^0)$,   metrics $g^0, g^2$ are Ricci-flat and $g^1, g^3$
are flat. (Since $(M_3,g^3)$ is 2-dimensional Ricci-flat space, it
is flat.) The metrics $g^i$, $i = 0,1,2$ have Euclidean signatures
and the metric $g^3$ has the signature $(-,+)$. The brane sets are
$I_1 = \{1,3 \}$ and $I_2 = \{2,3 \}$ for $M2$ and $M5$ branes,
respectively.  We put here $M_1 = \R$.

Let us introduce  the following set of $\Gamma$-matrices

 \bear{4.4}
  (\hat{\Gamma}^A) =
  (\hat{\Gamma}^{a_{0}}_{(0)} \otimes 1 \otimes \1_4 \otimes
  \1_2, \\  \nn
   \hat{\Gamma}_{(0)} \otimes 1 \otimes \hat{\Gamma}_{(2)} \otimes
   \hat{\Gamma}_{(3)}, \\  \nn
   \hat{\Gamma}_{(0)} \otimes 1 \otimes \hat{\Gamma}_{(2)}^{a_{2}} \otimes
   \1_2, \\  \nn
   \hat{\Gamma}_{(0)} \otimes 1 \otimes \hat{\Gamma}_{(2)} \otimes
   \hat{\Gamma}_{(3)}^{a_{3}}). \\  \nn
  \ear

Here
  \beq{4.4a}
 \hat{\Gamma}_{(0)} = \hat{\Gamma}_{(0)}^{1_0}
 \hat{\Gamma}_{(0)}^{2_0} \hat{\Gamma}_{(0)}^{3_0}
 \hat{\Gamma}_{(0)}^{4_0}, \quad
 \hat{\Gamma}_{(2)} =
 \hat{\Gamma}_{(2)}^{1_2} \hat{\Gamma}_{(2)}^{2_2}
 \hat{\Gamma}_{(2)}^{3_2} \hat{\Gamma}_{(2)}^{4_2}, \quad
 \hat{\Gamma}_{(3)} = \hat{\Gamma}_{(3)}^{1_3}
 \hat{\Gamma}_{(3)}^{2_3}
  \eeq
  obey
 \beq{4.4b}
 (\hat{\Gamma}_{(0)})^2 = (\hat{\Gamma}_{(2)})^2 = \1_4,
 \qquad  (\hat{\Gamma}_{(0)})^2  = \1_2.
 \eeq

  Let
   \beq{4.6}
   \eta =  \eta_0(x) \otimes \eta_1(y_1)\otimes \eta_2(y_2)\otimes \eta_3(y_3),
   \eeq
 where $\eta_0 =   \eta_0(x)$ is $4$-component spinor on $M_0$,
       $\eta_1 = \eta_1(y_1)$ is $1$-component spinor on  $M_1$,
       $\eta_2 = \eta_2(y_2)$ is $4$-component spinor on $M_2$,
       and $\eta_3 = \eta_3(y_3)$ is $2$-component spinor on $M_3$.

   It follows from  (\ref{0.38}),  (\ref{4.4})  and (\ref{4.4b}) that
    \bear{4.7a}
     \bar{D}_{m_0}^{(0)} \eta  =  (D_{m_0}^{(0)} \eta_0) \otimes
      \eta_1 \otimes \eta_2 \otimes  \eta_3, \quad
     \bar{D}_{m_1}^{(1)} \eta =  \eta_0 \otimes (D_{m_1}^{(1)}
     \eta_1) \otimes \eta_2 \otimes  \eta_3  , \\ \nn
     \bar{D}_{m_2}^{(2)} \eta =  \eta_0 \otimes \eta_1 \otimes (D_{m_2}^{(2)}
     \eta_2) \otimes  \eta_3,  \quad
     \bar{D}_{m_3}^{(3)} \eta = \eta_0 \otimes \eta_1 \otimes \eta_2 \otimes (D_{m_3}^{(3)}
     \eta_3),
     \ear
  where $D^{(i)}_{m_i}$ are covariant (spinorial) derivative (\ref{0.38a})
  corresponding the manifold  $M_i$, $i = 0,1,2,3$. Here $D^{(1)}_{m_1} = \p_{m_1}$.

  The operator (\ref{2.14e}) corresponding to $M2$-brane reads
   \beq{4.8a}
     \hat{\Gamma}_{[s]} =  \hat{\Gamma}^{1_1}  \hat{\Gamma}^{1_3} \hat{\Gamma}^{2_3} =
    \hat{\Gamma}_{(0)} \otimes 1 \otimes  \hat{\Gamma}_{(2)} \otimes  \1_2,
     \eeq
    $s = I_1$ and the operator (\ref{2.14m}) corresponding to
    $M5$-brane has the following form
    \beq{4.8b}
    \hat{\Gamma}_{[s]} =  \hat{\Gamma}^{1_0} \hat{\Gamma}^{2_0} \hat{\Gamma}^{3_0}
                  \hat{\Gamma}^{4_0}  \hat{\Gamma}^{1_1} =
      \1_4  \otimes 1 \otimes  \hat{\Gamma}_{(2)} \otimes \hat{\Gamma}_{(3)} ,
     \eeq
    $s = I_2$.    Then  the restrictions  (\ref{2.16b})
     are satisfied if
     \beq{4.8c}
     \hat{\Gamma}_{(j)} \eta_j  = c_{(j)} \eta_j, \qquad c_{(j)}^2 = 1,
     \eeq
    $j = 0,2,3$, and
    \beq{4.8d}
     c_{(0)} c_{(2)} = c_1, \qquad  c_{(2)} c_{(3)} = c_2.
     \eeq

 Using the Proposition 1 we obtain the following
 solution to  equations (\ref{1.2})
 corresponding to the field configuration from
(\ref{4.2}), (\ref{4.3})
 \beq{4.9}
 \varepsilon  = H_1^{-1/6} H_2^{-1/12}
 \eta_0(x) \otimes \eta_1 \otimes \eta_2(y_2)\otimes \eta_3(y_3).
 \eeq
 Here  $\eta_i$, $i = 0,2,3$, are parallel  chiral spinors
 defined on $M_i$, respectively ($D^{(i)}_{m_i} \eta_i = 0$),
  obeying  (\ref{4.8c}) and (\ref{4.8d}); $\eta_1$ is constant
  (1-dimensional spinor).

  It follows from (\ref{4.8d}) that either
  i) $c_{(2)} = 1$, $c_{(0)}  = c_1$, $c_{(3)}  = c_2$,
  or
  ii) $c_{(2)} =  - 1$, $c_{(0)} = - c_1$, $c_{(3)}  =  - c_2$.

  Thus, the number of linear independent solutions
  given by (\ref{4.9}) is

  \beq{4.9a}
   N  =  32 {\cal N} =  n_0( c_1)  n_2(1) n_3(c_2)
          +  n_0(- c_1) n_2(-1)  n_3(- c_2).
   \eeq
 Here $n_j (c_{(j)})$ is the number of chiral parallel spinors on $M_j$
 obeying  (\ref{4.8c}),  $j = 0,2,3$.

  {\bf $\R^{1,1}$-intersection.} Let $M_3 = \R^2$,
 $g^3 = - d y^1_3  \otimes d y^1_3 + d y^2_3 \otimes d y^2_3$.
 Thus we deal with 2-dimensional pseudo-Euclidean space $\R^{1,1}$.
 In this case   we get $n_3 (c) = 1$ and the number of unbroken SUSY is at
 least

   \beq{4.11}
   { \cal N } = \frac{1}{32}
        \sum_{c = \pm 1} n_0( c c_1) n_2(c ).
    \eeq

  For flat $M_0 = M_2 = \R^4$ we have $n_0(c) =  n_2(c) = 2$ and
  hence ${\cal N} = 1/4$.

   {\bf Example: one $K3$ factor-space.} Let $M_0 = \R^4$ and $M_2 = K3$. We get
   from (\ref{4.11}): ${ \cal N } = 1/8$. The same number of fractional SUSY
    will be obtained for  $M_0 = K3$ and $M_2 = \R^4$.

   {\bf Example: two $K3$ factor-spaces.} Let $M_0 =  M_2 = K3$.
    We put for chiral numbers   $(n_i(+1), n_i(-1)) = (2, 0)$, $i =
    0,2$. Then we get ${\cal N} = 1/8$ for $c_1 = + 1$ and ${\cal N} = 0$
    for $c_1 = - 1$. It should be noted that here the number of fractional SUSY
    depends only upon the sign-factor of electric $M2$-brane $c_1$. It is
    independent upon the  sign-factor of electric $M5$-brane
    $c_2$. The change of the orientation of one of the manifolds, say $M_2$,
    i.e. when $M_2 = K3$, with $(n_2(+1), n_2(-1)) = (0,2)$
    will lead to ${\cal N} = 1/8$ for $c_1 = - 1$ and ${\cal N} = 0$
    for $c_1 = + 1$.

   \subsection{$M2 \cap M2$-branes}

 Consider a solution  with two intersecting $M2$-branes defined on
the manifold (\ref{4.1})

 $$M_{0}  \times M_{1}  \times  M_{2} \times M_{3},$$
 with $d_0 = 6$, $d_1 = d_2 = 2$ and  $d_3 = 1$.

 The solution reads
 \bear{5.2}
 g= H_1^{1/3} H_2^{1/3}
 \{ \hat{g}^0  + H_1^{-1} \hat{g}^1 +  H_2^{-1} \hat{g}^2
 + H_1^{-1} H_2^{-1} \hat{g}^3 \},
 \\ \label{5.3}
 F = c_1 d H^{-1}_1 \wedge  \hat{\tau}_1 \wedge \hat{\tau}_3 +
     c_2 d H^{-1}_2 \wedge  \hat{\tau}_2 \wedge \hat{\tau}_3                                             ,
 \ear
where $c^2_1 = c^2_2 = 1$; $H_1$, $H_2$ are harmonic functions on
6-dimensional Ricci-flat Riemann manifold $(M_0,g^0)$,  metrics
$g^i$, $i = 1,2$, are 2-dimensional flat metrics of Euclidean
signature $(+,+)$.  The brane sets are $I_1 = \{1,3 \}$ and $I_2 =
\{2,3 \}$. We put here $M_3 = \R$ and $g^3 = - dt \otimes dt$
($\tau_3 = dt$).

Let us consider  the following set of $\Gamma$-matrices
 \bear{5.4}
 (\hat{\Gamma}^A) =
  (\hat{\Gamma}^{a_{0}}_{(0)}  \otimes \1_2 \otimes
  \1_2 \otimes 1, \\  \nn
  i \hat{\Gamma}_{(0)}  \otimes \hat{\Gamma}_{(1)}^{a_{1}} \otimes
   \1_{2} \otimes 1, \\  \nn
   \hat{\Gamma}_{(0)}  \otimes \hat{\Gamma}_{(1)} \otimes
    \hat{\Gamma}_{(2)}^{a_{2}} \otimes 1, \\  \nn
   \hat{\Gamma}_{(0)}  \otimes \hat{\Gamma}_{(1)} \otimes
   \hat{\Gamma}_{(2)} \otimes 1). \\  \nn
  \ear

Here
  \beq{5.4a}
 \hat{\Gamma}_{(0)} = \hat{\Gamma}_{(0)}^{1_0} \dots
 \hat{\Gamma}_{(0)}^{6_0}, \quad
 \hat{\Gamma}_{(1)} =
 \hat{\Gamma}_{(1)}^{1_1} \hat{\Gamma}_{(1)}^{2_1}, \quad
 \hat{\Gamma}_{(2)} =
 \hat{\Gamma}_{(2)}^{1_2} \hat{\Gamma}_{(2)}^{2_2}
  \eeq
  obey
 \beq{5.4b}
 (\hat{\Gamma}_{(0)})^2 = - \1_8, \qquad
 (\hat{\Gamma}_{(1)})^2 = (\hat{\Gamma}_{(2)})^2 = - \1_2.
 \eeq

  As in the previous case we consider the ansatz (\ref{4.6})
   $\eta =  \eta_0(x) \otimes \eta_1(y_1)\otimes \eta_2(y_2)\otimes \eta_3(y_3)$,
    where $\eta_0 =   \eta_0(x)$ is $8$-component spinor on $M_0$,
       $\eta_1 = \eta_1(y_1)$ is $2$-component spinor  on  $M_1$,
       $\eta_2 = \eta_2(y_2)$ is $2$-component spinor  on $M_2$,
       and $\eta_3 = \eta_3(y_3)$ is $1$-component spinor on
       $M_3$.

   Due to  (\ref{0.38}),  (\ref{5.4}),   and (\ref{5.4b})
   the relations  (\ref{4.7a}) are satisfied identically. Here
     $D^{(3)}_{m_3} = \p_{m_3}$.

  The  operators (\ref{2.14e}) corresponding to $M2$-branes read
   \beq{5.8a}
     \hat{\Gamma}_{[s]} =  \hat{\Gamma}^{1_1}  \hat{\Gamma}^{2_1} \hat{\Gamma}^{1_3} =
   - \hat{\Gamma}_{(0)}  \otimes   \1_2 \otimes \hat{\Gamma}_{(2)}  \otimes 1,
     \eeq
    for $s =  I_1$ and
    \beq{5.8b}
     \hat{\Gamma}_{[s]} =  \hat{\Gamma}^{1_2} \hat{\Gamma}^{2_2} \hat{\Gamma}^{1_3}
            =   - \hat{\Gamma}_{(0)} \otimes \hat{\Gamma}_{(1)} \otimes \1_2 \otimes 1,
     \eeq
    for $s = I_2$.    Then  the chirality restrictions (\ref{2.16b})
     are satisfied if
     \beq{5.8c}
     \hat{\Gamma}_{(j)} \eta_j  = c_{(j)} \eta_j, \qquad  c_{(j)}^2 =  -1,
     \eeq
   $j = 0,1,2$, and
    \beq{5.8d}
    - c_{(0)} c_{(2)} = c_1, \qquad  - c_{(0)} c_{(1)} = c_2.
     \eeq

 Using the Proposition 1 we get the following
 solution to SUSY equations (\ref{1.2})
 corresponding to the field configuration from
 (\ref{5.2}), (\ref{5.3})
 \beq{5.9}
 \varepsilon  = H_1^{-1/6} H_2^{-1/6}
 \eta_0(x) \otimes \eta_1(y_1) \otimes \eta_2(y_2) \otimes \eta_3.
 \eeq
 Here  $\eta_i$, $i = 0,1,2$, are chiral parallel spinors
 defined on $M_i$, respectively: $D^{(i)}_{m_i} \eta_i = 0$,
  obeying  (\ref{5.8c}) and (\ref{5.8d}); $\eta_3$ is constant
  ($1$-component spinor).

  It follows from (\ref{5.8d}) that either
  i) $c_{(0)} = i$, $c_{(2)}  = i c_1$, $c_{(1)}  = i c_2$,
  or
  ii) $c_{(2)} =  - i$, $c_{(2)}  = - i c_1$, $c_{(1)}  =  - i c_2$.

  Thus, the number of linear independent solutions
  given by (\ref{5.9}) is
  \beq{5.9a}
   N  =   32 {\cal N} =  n_0(+i) n_1(  i c_2)  n_2(  i c_1)
              +  n_0(-i) n_1(- i c_2)  n_2(- i c_1),
   \eeq
 where $n_j (c_{(j)})$ is the number of chiral parallel spinors on $M_j$
 obeying  (\ref{5.8c}),  $j = 0,1,2$.

 We remind that here  $M_1$ and $M_2$ are 2-dimensional flat spaces of Euclidean signature.

  {\bf Case $M_1 = M_2 = \R^2$.} Let $M_1 = M_2 = \R^2$. Then
  due to  $n_1 (c) = n_2 (c) = 1$  the number of unbroken SUSY is at least

   \beq{5.11}
   {\cal N}  = \frac{1}{32} n_0.
  \eeq

  Here ${\cal N}$ does not depend upon brane signs $c_s$.
  For flat $M_0 = \R^6$ we have $n_0 = 8$ and
  hence ${\cal N} = 1/4$.

  {\bf Example: $M_0 = CY_3$.}
  Let $M_0 = CY_3$ be 6-dimensional Calabi-Yau manifold (3-fold) of holonomy $SU(3)$.
  Then $n_0 (i) = n_0 (-i) = 1$,  $n_0 = 2$ and hence ${\cal N} =
  1/16$.

  {\bf Example: $M_0 = \R^2 \times K3$.}
  For $M_0 = \R^2 \times K3$  we get $n_0 (i) = n_0 (-i) = 2$,  $n_0 =4$
  and consequently ${\cal N} = 1/8$.

   \subsection{$M5 \cap M5$-branes}

Now we deal with $M5 \cap M5$-solution defined on the manifold
(\ref{4.1})

$$M_{0}  \times M_{1}  \times M_{2} \times M_{3},$$

 with $d_0 = 3 $, $d_1 = d_2 = 2$ and $d_3 = 4$.

 The solution reads
 \bear{6.2}
 g= H_1^{2/3} H_2^{2/3}
 \{ \hat{g}^0  + H_1^{-1} \hat{g}^1 +  H_2^{-1} \hat{g}^2
     + H_1^{-1} H_2^{-1} \hat{g}^3 \},
 \\ \label{6.3}
 F = c_1   (*_0 d H_1) \wedge \hat{\tau}_2 +
     c_2   (*_0 d H_2) \wedge \hat{\tau}_1,
 \ear
where $c^2_1 = c^2_2 = 1$; $H_1$, $H_2$ are harmonic functions on
$(M_0,g^0)$,   metrics $g^i$, $i = 0,1,2,3$, are Ricci-flat ( the
first three metrics are flat). The metrics $g^i$, $i = 0,1,2$,
have Euclidean signatures and the metric $g^3$ has the signature
 $(-,+,+,+)$. The brane sets are $I_1 = \{1,3 \}$ and
 $I_2 = \{2,3 \}$.

We choose  the following  $\Gamma$-matrices
 \bear{6.4}
  (\hat{\Gamma}^A) =
  (i \hat{\Gamma}^{a_{0}}_{(0)}  \otimes \hat{\Gamma}_{(1)} \otimes
   \hat{\Gamma}_{(2)} \otimes \hat{\Gamma}_{(3)}, \\  \nn
   \1_2  \otimes \hat{\Gamma}_{(1)}^{a_{1}} \otimes
   \hat{\Gamma}_{(2)} \otimes \hat{\Gamma}_{(3)}, \\  \nn
    i \1_2  \otimes   \1_2 \otimes
    \hat{\Gamma}_{(2)}^{a_{2}} \otimes \hat{\Gamma}_{(3)}, \\  \nn
    \1_2  \otimes   \1_2 \otimes  \1_2 \otimes \hat{\Gamma}_{(3)}^{a_{3}}), \\  \nn
  \ear
where
  \beq{6.4a}
   \hat{\Gamma}_{(0)} = \hat{\Gamma}_{(0)}^{1_0} \hat{\Gamma}_{(0)}^{2_0}
   \hat{\Gamma}_{(0)}^{3_0}, \quad
 \hat{\Gamma}_{(1)} =
 \hat{\Gamma}_{(1)}^{1_1} \hat{\Gamma}_{(1)}^{2_1},
 \quad
 \hat{\Gamma}_{(2)} =
 \hat{\Gamma}_{(2)}^{1_2} \hat{\Gamma}_{(2)}^{2_2},
 \quad
 \hat{\Gamma}_{(3)} =
 \hat{\Gamma}_{(3)}^{1_3} \hat{\Gamma}_{(3)}^{2_3}
 \hat{\Gamma}_{(3)}^{3_3} \hat{\Gamma}_{(3)}^{4_3},
  \eeq
  obey
 \beq{6.4b}
 (\hat{\Gamma}_{(0)})^2 =
 (\hat{\Gamma}_{(1)})^2 = (\hat{\Gamma}_{(2)})^2 =  - \1_2,
 \quad (\hat{\Gamma}_{(3)})^2 =  - \1_4.
 \eeq

 We put $(\hat{\Gamma}^{a_{0}}_{(0)}) = (\sigma_1, \sigma_2, \sigma_3)$
 and hence
   \beq{6.4c}  \hat{\Gamma}_{(0)} = i \1_2. \eeq

  Here we consider the decomposition (\ref{4.6})
   $\eta =  \eta_0(x) \otimes \eta_1(y_1)\otimes \eta_2(y_2)\otimes \eta_3(y_3)$,
    where $\eta_0 =   \eta_0(x)$ is $2$-component spinor on $M_0$,
       $\eta_1 = \eta_1(y_1)$ is $2$-component spinor on  $M_1$,
       $\eta_2 = \eta_2(y_2)$ is $2$-component spinor on $M_2$,
       and $\eta_3 = \eta_3(y_1)$ is $4$-component spinor on $M_3$.

   Due to  (\ref{0.38}),  (\ref{6.4}) and (\ref{6.4b})
    relations  (\ref{4.7a}) are also satisfied in this case.

     The  operators (\ref{2.14m}) corresponding to $M5$-branes read
   \beq{6.8a}
     \hat{\Gamma}_{[s]} = \hat{\Gamma}^{1_0}  \hat{\Gamma}^{2_0} \hat{\Gamma}^{3_0}
                  \hat{\Gamma}^{1_2} \hat{\Gamma}^{2_2}
     =    \1_2 \otimes \hat{\Gamma}_{(1)}  \otimes   \1_2 \otimes \hat{\Gamma}_{(3)},
     \eeq
    for $s =  I_1$ and
    \beq{6.8b}
     \hat{\Gamma}_{[s]} = \hat{\Gamma}^{1_0}  \hat{\Gamma}^{2_0} \hat{\Gamma}^{3_0}
                  \hat{\Gamma}^{1_1} \hat{\Gamma}^{2_1}
     =    \1_2  \otimes   \1_2 \otimes \hat{\Gamma}_{(2)}  \otimes \hat{\Gamma}_{(3)},
     \eeq
    for $s = I_2$.    Then  the restrictions (\ref{2.16b})
     are satisfied if
     \beq{6.8c}
     \hat{\Gamma}_{(j)} \eta_j  = c_{(j)} \eta_j, \qquad  c_{(j)}^2 =  -1,
     \eeq
  $j = 1,2,3$, and
    \beq{6.8d}
     c_{(1)} c_{(3)} = c_1, \qquad   c_{(2)} c_{(3)} = c_2.
     \eeq

According to   Proposition 1 we get the following  solution to
equations (\ref{1.2})  corresponding to the field configuration
from (\ref{6.2}), (\ref{6.3})
 \beq{6.9}
 \varepsilon  = H_1^{-1/12} H_2^{-1/12}
 \eta_0(x) \otimes \eta_1(y_1) \otimes \eta_2(y_2) \otimes \eta_3(y_3).
 \eeq
 Here $\eta_0$ is parallel spinor on $M_0$ and
 $\eta_i$, $i = 1,2,3$, are chiral parallel spinors
 defined on $M_i$, respectively ($D^{(i)}_{m_i} \eta_i = 0$),
  obeying  (\ref{6.8c}) and (\ref{6.8d}).

  It follows from (\ref{6.8d}) that either
  i)  $c_{(3)} = i$, $c_{(1)}  = - i c_1$, $c_{(2)}  = - i c_2$,
  or
  ii) $c_{(3)} =  - i$, $c_{(1)}  =  i c_1$, $c_{(2)}  =   i c_2$.

  Thus, the number of linear independent solutions
  given by (\ref{6.9}) is
  \beq{6.9a}
   N  =  32 {\cal N} =  n_0  n_1(-i c_1)  n_2(-i c_2) n_3(+i)
             +  n_0  n_1( i c_1)  n_2( i c_2) n_3(-i) .
   \eeq

 Here $n_j (c_{(j)})$ is the number of chiral parallel spinors on $M_j$
 (see (\ref{6.8c})),  $j = 1,2,3$, and $n_0$ is the number of
  parallel spinors on $M_0$.

  {\bf  Case $M_1 = M_2 = \R^2$.}

  Let $M_1 = M_2 = \R^2$. We get
    $n_1 (c) = n_2 (c) = 1$ and hence

   \beq{6.11}
    { \cal N} = \frac{1}{32} n_0  n_3.
  \eeq

  For flat factor-spaces with $M_0 = \R^3$ (and Euclidean metric) and
  $M_3 = \R^4$ (and Minkowski metric of $\R^{1,3}$) we get   ${\cal N} = 1/4$.

  {\bf Example: 4-dimensional $pp$-wave metric. }
  Let $(M_3,g^3)$ be a 4-dimensional Ricci-flat $pp$-wave solution
  from \cite{FOF}  with the holonomy group $H = \R^2$. Then we
  get $n_3 (1) = n_3 (-1) = 1$, $n_3 = 2$  ( see \cite{Bohle}, p.
  53)   and hence ${\cal N} =   n_0/16$. For $M_0 = \R^3$ we are led to
  ${\cal N} =   1/8$.

 \subsection{$M2 \cap M2 \cap M2$-branes}

The last example is related to the solution with  three
intersecting $M2$-branes defined on the manifold

\beq{7.1}
 M_{0}  \times M_{1}  \times  M_{2} \times M_{3} \times M_{4},
 \eeq
 with $d_0 = 4$, $d_1 = d_2 = d_3 = 2$ and  $d_4 = 1$.

 The solution reads
 \bear{7.2}
 g= H_1^{1/3} H_2^{1/3} H_3^{1/3}
 \{ \hat{g}^0  + H_1^{-1} \hat{g}^1 +  H_2^{-1} \hat{g}^2
   +  H_3^{-1} \hat{g}^3 + H_1^{-1} H_2^{-1} H_3^{-1} \hat{g}^4 \},
 \\ \label{7.3}
 F =  c_1 d H^{-1}_1 \wedge  \hat{\tau}_1 \wedge \hat{\tau}_4 +
       c_2 d H^{-1}_2 \wedge  \hat{\tau}_2 \wedge \hat{\tau}_4
      +  c_3 d H^{-1}_3 \wedge  \hat{\tau}_3 \wedge \hat{\tau}_4                                             ,
 \ear

where $c^2_1 = c^2_2 = c^2_3 =1$; $H_1$, $H_2$, $H_3$ are harmonic
functions on $(M_0,g^0)$,   metrics $g^i$, $i = 0,1,2,3,4$, are
Ricci-flat (the last four metrics are flat). The metrics $g^i$, $i
= 0,1,2,3$, have Euclidean signatures and the metric $g^4$ has the
signature $(-)$. Here we put $M_4 = \R$, $g^4 = - dt \otimes dt$
($\tau_4 = dt$). The brane sets are $I_1 = \{1,4 \}$, $I_2 = \{2,4
\}$ and $I_3 = \{3,4 \}$.

Let us consider the following set of $\Gamma$-matrices
 \bear{7.4}
  (\hat{\Gamma}^A) =
   (\hat{\Gamma}^{a_{0}}_{(0)}  \otimes \1_2
   \otimes  \1_2 \otimes  \1_2 \otimes 1, \\  \nn
   \hat{\Gamma}_{(0)}  \otimes \hat{\Gamma}_{(1)}^{a_{1}}
   \otimes  \1_2 \otimes  \1_{2} \otimes 1, \\  \nn
  i \hat{\Gamma}_{(0)}  \otimes \hat{\Gamma}_{(1)} \otimes
   \hat{\Gamma}_{(2)}^{a_{2}} \otimes  \1_2 \otimes 1, \\  \nn
   \hat{\Gamma}_{(0)}  \otimes \hat{\Gamma}_{(1)} \otimes
   \hat{\Gamma}_{(2)} \otimes \hat{\Gamma}_{(3)}^{a_{3}} \otimes 1, \\  \nn
   \hat{\Gamma}_{(0)}  \otimes \hat{\Gamma}_{(1)} \otimes
   \hat{\Gamma}_{(2)} \otimes \hat{\Gamma}_{(3)} \otimes 1),
  \ear
 where
  \beq{7.4a}
 \hat{\Gamma}_{(0)} = \hat{\Gamma}_{(0)}^{1_0} \dots
 \hat{\Gamma}_{(0)}^{4_0}, \quad
 \hat{\Gamma}_{(i)} =
 \hat{\Gamma}_{(i)}^{1_i} \hat{\Gamma}_{(i)}^{2_i},
  \eeq
   obey
 \beq{7.4b}
 (\hat{\Gamma}_{(0)})^2 =  \1_4, \qquad
 (\hat{\Gamma}_{(i)})^2 = - \1_2,
 \eeq
 $i = 1,2,3$.

  Here  we put
  \beq{7.6}
   \eta =  \eta_0(x) \otimes \eta_1(y_1)\otimes \eta_2(y_2)\otimes \eta_3(y_3)
   \otimes \eta_4(y_4),
   \eeq
  where $\eta_0 =   \eta_0(x)$ is $4$-component spinor  on $M_0$,
  $\eta_i = \eta_i(y_i)$ is $2$-component spinor on $M_i$,
  $i = 1,2,3$,  and $\eta_4 = \eta_4(y_4)$ is $1$-component spinor on   $M_4$.

    Due to  (\ref{0.38}),  (\ref{7.4}) and (\ref{7.4b})
   the following relations take place
     \bear{7.7a}
     \bar{D}_{m_0}^{(0)} \eta  =  (D_{m_0}^{(0)} \eta_0) \otimes
      \eta_1 \otimes \eta_2 \otimes  \eta_3 \otimes  \eta_4, \quad
     \bar{D}_{m_1}^{(1)} \eta =  \eta_0 \otimes (D_{m_1}^{(1)}
     \eta_1) \otimes \eta_2 \otimes  \eta_3 \otimes  \eta_4 , \\ \nn
     \bar{D}_{m_2}^{(2)} \eta =  \eta_0 \otimes \eta_1 \otimes (D_{m_2}^{(2)}
     \eta_2) \otimes  \eta_3 \otimes  \eta_4,  \quad
     \bar{D}_{m_3}^{(3)} \eta =  \eta_0 \otimes \eta_1 \otimes  \eta_2 \otimes (D_{m_3}^{(3)}
     \eta_3) \otimes  \eta_4, \\ \nn
     \bar{D}_{m_4}^{(4)} \eta = \eta_0 \otimes \eta_1 \otimes
     \eta_2 \otimes \eta_3  \otimes (D_{m_4}^{(4)}  \eta_4),
     \ear
   where $D^{(i)}_{m_i}$   are covariant  derivative
   corresponding to  $M_i$, $i = 0,1,2,3$. Here $D^{(4)}_{m_4} = \p_{m_4}$.

  The  operators (\ref{2.14e}) corresponding to $M2$-branes read
   \beq{7.8a}
     \hat{\Gamma}_{[s]} =  \hat{\Gamma}^{1_1}  \hat{\Gamma}^{2_1} \hat{\Gamma}^{1_4} =
     - \hat{\Gamma}_{(0)}  \otimes   \1_2 \otimes \hat{\Gamma}_{(2)}
     \otimes \hat{\Gamma}_{(3)} \otimes 1
     \eeq
    for $s =  I_1$,
    \beq{7.8b}
     \hat{\Gamma}_{[s]} =  \hat{\Gamma}^{1_2} \hat{\Gamma}^{2_2} \hat{\Gamma}^{1_4}
     =   - \hat{\Gamma}_{(0)} \otimes \hat{\Gamma}_{(1)} \otimes \1_2
           \otimes \hat{\Gamma}_{(3)} \otimes 1
     \eeq
    for $s = I_2$ and
    \beq{7.8c}
     \hat{\Gamma}_{[s]} = \hat{\Gamma}^{1_3} \hat{\Gamma}^{2_3} \hat{\Gamma}^{1_4}
     =   - \hat{\Gamma}_{(0)} \otimes \hat{\Gamma}_{(1)} \otimes \hat{\Gamma}_{(2)}
            \otimes \1_2  \otimes 1
     \eeq
      for $s = I_3$.

     We put
     \bear{7.8ca}
     \hat{\Gamma}_{(0)} \eta_0  = c_{(0)} \eta_0, \qquad  c_{(0)}^2 =  1,
      \\ \label{7.8cb}
     \hat{\Gamma}_{(j)} \eta_j  = c_{(j)} \eta_j, \qquad  c_{(j)}^2 = -1,
     \ear
   $j = 1,2,3$. Then  the chirality restrictions (\ref{2.16b})
     are satisfied if
    \beq{7.8d}
    - c_{(0)} c_{(2)} c_{(3)} = c_1, \qquad
     - c_{(0)} c_{(1)} c_{(3)} = c_2,
     \qquad  - c_{(0)} c_{(1)} c_{(2)} = c_3.
     \eeq

 Due to  Proposition 1 we obtain the following
 solution to SUSY equations (\ref{1.2})
 corresponding to the field configuration from
 (\ref{7.2}), (\ref{7.3})
 \beq{7.9}
 \varepsilon  = H_1^{-1/6} H_2^{-1/6} H_3^{-1/6}
 \eta_0(x) \otimes \eta_1(y_1) \otimes \eta_2(y_2) \otimes \eta_3(y_3) \otimes
 \eta_4.
 \eeq
 Here  $\eta_i$, $i = 0,1,2,3$, are chiral parallel spinors
 defined on $M_i$, respectively ($D^{(i)}_{m_i} \eta_i = 0$),
  obeying  (\ref{7.8ca}), (\ref{7.8cb}) and (\ref{7.8d}); $\eta_4$ is constant.

  Equations (\ref{7.8d}) have the following  solutions
  \beq{7.9aa}
   c_{(0)} = c_1 c_2 c_3, \qquad  c_{(j)}  = \pm  i c_j,
   \eeq
   $j = 1,2,3$.

  Thus, the number of linear independent solutions
  given by (\ref{7.9})  and (\ref{7.9aa})  is

  \beq{7.9a}
   N  = 32 {\cal N} =  n_0(c_1 c_2 c_3)
            \sum_{c = \pm 1} n_1( i c c_1) n_2( i c
            c_2) n_3(  i c c_3),
   \eeq
 where $n_j (c_{(j)})$ is the number of chiral parallel spinors on
 $M_j$, $j = 0,1,2,3$; see (\ref{7.8ca}) and (\ref{7.8cb}).

 Here 2-dimensional  manifolds $M_1$, $M_2$ and $M_3$ are
 flat.

{\bf  Case $M_1 = M_2 = M_3 = \R^2$.}
 Let  $M_1 = M_2 = M_3 = \R^2$.  Then all $n_j (ic) = 1$,
 $j = 1,2,3$, $c = \pm 1$, and hence we get from (\ref{7.9a})
  we get
   \beq{7.11}
   {\cal N}  = \frac{1}{16} n_0( c_1 c_2 c_3).
  \eeq

  For  $M_0 = \R^4$ we have $n_0(c) = 2$ and   hence ${\cal N } = 1/8$ for all values
  of $c_i$, $i = 1,2,3$.

  {\bf Example: $M_0 = K3$.}
  Let $M_0 = K3$ with $n_0(1) = 2$ and $n_0(-1) = 0$. Then we
  get ${\cal N } = 1/8$ if  $c_1 c_2 c_3 = 1$  and
  ${\cal N } = 0$ if  $c_1 c_2 c_3 = - 1$.

\section{Conclusions and discussions}

In this paper  we have considered the ``Killing-like''  SUSY
equations in $D=11$  supergravity  for non-marginal  $M$-brane
solutions defined on the product of Ricci-flat manifolds $M_{0}
\times M_{1} \times \ldots \times M_{n}$ \cite{IM0,IM01}.

By proving the Proposition 1 we have found the solutions to these
``Killing-like''   equations which are defined up to  the
solutions to first-order differential equations:
$\bar{D}_{m_l}^{(l)} \eta = 0$, $l = 0, \dots, n$, with brane
``chirality'' conditions $\hat{\Gamma}_{[s]} \eta = c_s \eta$
imposed. The operators $\bar{D}_{m_l}^{(l)}$ after a proper choice
of $\Gamma$-matrices written in the tensor product form acts on
$32$-component spinor  $\eta = \eta_0 \otimes ... \otimes \eta_n$
as following $\bar{D}_{m_l}^{(l)} \eta =  ... \otimes \eta_{l-1}
\otimes D_{m_l}^{(l)} \eta_l  \otimes \eta_{l+1} \otimes ...$,
where $D_{m_l}^{(l)}$ is the spinorial covariant derivative
corresponding to the manifold $M_l$. Thus, the problem of finding
the solutions to SUSY equations is reduced here to  the search of
chiral parallel spinors on factor-spaces $M_l$ and to the
(technical) task of finding suitable sets of  $\Gamma$-matrices
written in the tensor product form corresponding to the product
manifold  $M_{0} \times M_{1} \times \ldots \times M_{n}$.

This program was successfully fulfilled here for the following
brane configurations: $M2$, $M5$, $M2 \cap M2$, $M2 \cap M5$, $M5
\cap M5$ and $M2 \cap M2 \cap M2$ and formulae for fractional
numbers of unbroken supersymmetries ${\cal N}$ were obtained. (The
formulae for $M2$-, $M5$-brane solutions were  obtained earlier in
\cite{Iv-00}.) Here we have considered certain examples of
partially supersymmetric configurations for various factor-spaces
$M_i$.

In the next publications we plan to complete the list of all
partially supersymmetric non-marginal $M$-brane configurations on
product of Ricci-flat factor-spaces along a line as it was done by
E. Bergshoeff {\it et al} \cite{BREJS} for $M_i = \R^{k_i}$, $i =
0, \dots, n$.  The results of this paper may be used in studies of
partially supersymmetric solutions defined on product of
Ricci-flat manifolds which take place in $IIA$, $IIB$ and other
low-dimensional supergravities. One can extend this formalism to
the so-called ``pseudo-supersymmetric'' p-brane solutions,
suggested recently in \cite{LuW}.

Another topic of interest may be in analyzing of special solutions
with moduli functions $H_s = C_s +  Q_s/ r^{d_0 -2}$  ($C_s \geq
0$, $Q_s > 0$) which are defined on  $(M_0, g^0)$ being a cone
over certain Einstein space   $(X,h)$, i.e. $g^0 = dr \otimes dr +
r^2 h$ ($X = S^{d_0 - 1}$ for $M_0 = \R^{d_0}$). In the
``near-horizon'' case $C_s = 0$ the fractional numbers of unbroken
SUSY might actually be larger (e.g. twice larger) then ``at
least'' numbers ${\cal N}$ obtained here for generic
$H_s$-functions. Here we will get Freund-Rubin-type solutions with
composite $M$-branes (see \cite{I-2} and references therein), e.g.
partially supersymmetric ones, that may of interest in a context
of AdS/CFT approach and its modifications.

\newpage

\begin{center}
{\bf Acknowledgments}
\end{center}

 This work was supported in part by  Russian Foundation for Basic Research
(Grant  Nr. 09-02-00677-a)  and by the  FTsP ``Nauchnie i
nauchno-pedagogicheskie kadry innovatsionnoy Rossii'' for the
years 2009-2013. The author is grateful to D.P. Sorokin, D.V.
Alekseevsky and J.M. Figueroa-O'Farrill for useful informing on
certain topics related to the subject of the paper.

 \renewcommand{\theequation}{\Alph{subsection}.\arabic{equation}}
 \renewcommand{\thesection}{}
 \renewcommand{\thesubsection}{\Alph{subsection}}
 \setcounter{section}{0}

     \section{Appendix}

     \subsection{The proof of relations (\ref{2.15a}) and (\ref{2.15b})}

      Here we prove relation (\ref{2.15a})

      $$E_{s} = - \sum_{i \in I_s} d_i \phi^i =  \ln H_s, \qquad s \in S_e,$$

       and relation (\ref{2.15b})      $$M_{s} = - \gamma (d_0 -2)   -
       \sum_{i \in \bar{I}_s} d_i \phi^i = -  \ln H_s, \qquad  s \in S_m.$$

      For the solution under consideration we have
            \beq{A.7}
            \gamma = \sum_{s \in S} \gamma_{s}, \qquad  \gamma_{s}= a_{s} \ln H_s
            \eeq
       and
           \beq{A.8}
            \phi^i = \sum_{s \in S} \phi^i_s, \qquad
            \phi^i_s =  (a_s - \frac{1}{2} \delta^i_{I_s}) \ln H_s,
            \eeq
       where $a_s = 1/6$ for $s \in S_e$ and $a_s = 1/3$ for $s \in
       S_m$.

       Relation (\ref{2.15a}) and (\ref{2.15b}) just follow from the
       identities
         \beq{A.9}
         E_{s}^{s'} = - \sum_{i \in I_s} d_i \phi^i_{s'} = \delta_{s}^{s'} \ln H_s, \qquad s \in
         S_e,
         \eeq
       and
       \beq{A.10}
        M_{s}^{s'} = - \gamma_{s'} (d_0 -2)   -
               \sum_{i \in \bar{I}_s} d_i \phi^i_{s'} = - \delta_{s}^{s'} \ln H_s, \qquad  s \in
               S_m,
       \eeq
       respectively.

       Let us  prove (\ref{A.9}). We get

         \beq{A.11}
         E_{s}^{s'} = {\cal E}_{s}^{s'} \ln H_s,
         \eeq
         where
         \beq{A.12}
         {\cal E}_{s}^{s'}=  - \sum_{i \in I_s} d_i (a_{s'} - \frac{1}{2}
         \delta^i_{I_{s'}}) = - a_{s'} d(I_s) + \frac{1}{2} d(I_s \cap
         I_{s'}).
         \eeq
         Here the following identity was used
         \beq{A.13}
          \sum_{i \in I} d_i  \delta^i_{J} =  d(I  \cap  J).
         \eeq
          (See (\ref{0.19}), (\ref{0.20a})). In what follows the relation $d(I_s) = 3$
         for $s \in S_e$ is used.

         In order to prove (\ref{A.9}) one should verify
         the equality
           \beq{A.14}
          {\cal E}_{s}^{s'} = \delta_{s}^{s'},
          \eeq
         for $s \in S_e$ and all $s'$.

         Let $s' \in
         S_m$,  then $a_{s'} = 1/3$ and $d(I_s  \cap  I_{s'}) = 2$ (due to intersection rules
         (\ref{2.8})). From (\ref{A.12}) we get
         ${\cal E}_{s}^{s'} = 0$ in agreement with (\ref{A.14}).

          Let $s' \in S_e$, then  $a_{s'} = 1/6$ and $d(I_s  \cap  I_{s'})
          = 3$ if $s = s'$ and $d(I_s  \cap  I_{s'})= 1$ if $s \neq
          s'$ (due to intersection rules (\ref{2.8})). Hence we
          obtain from (\ref{A.12}):
         ${\cal E}_{s}^{s'} = 1$ for $s = s'$ and
          ${\cal E}_{s}^{s'} = 0$ for $s \neq s'$
         in agreement with (\ref{A.14}). Thus,  relation (\ref{A.12})
         is proved and hence  relations (\ref{A.9}) and (\ref{2.15a})
         are also proved.

         Now we prove (\ref{A.10}). We get

         \beq{A.15}
         M_{s}^{s'} = {\cal M}_{s}^{s'} \ln H_s,
         \eeq
         where
         \bear{A.16}
         {\cal M}_{s}^{s'}=  - (d_0 -2)a_{s'} - \sum_{i \in \bar{I}_s} d_i (a_{s'} - \frac{1}{2}
         \delta^i_{I_{s'}}) \\ \nonumber
         = - (d_0 -2 + d(\bar{I}_s) )a_{s'}   + \frac{1}{2} d(\bar{I}_s \cap
          I_{s'}).
         \ear

         Using the definition of the dual set (\ref{2.7})  we
         have
           \beq{A.17}
           d(\bar{I}_s ) = 11 - d_0  - d(I_s )
           \eeq
           and
           \beq{A.18}
            d(\bar{I}_s \cap  I_{s'}) = d(I_{s'}) -  d(I_s \cap I_{s'}).
           \eeq
           In what follows the relation $d(I_s) = 6$ for $s \in S_m$ is used.

         In order to prove (\ref{A.10}) one should verify
         the equality
           \beq{A.19}
          {\cal M}_{s}^{s'} = - \delta_{s}^{s'},
           \eeq
         for $s \in S_m$ and all $s'$.

         Let $s' \in
         S_e$,  then $a_{s'} = 1/6$ and $d(I_{s'}) = 3$, $d(I_s  \cap  I_{s'}) = 2$ (due to intersection rules
         (\ref{2.8})). From (\ref{A.16}), (\ref{A.17}) and
         (\ref{A.18}) we obtain
         ${\cal M}_{s}^{s'} = 0$ in agreement with (\ref{A.19}).

          Let $s' \in S_m$, then  $a_{s'} = 1/3$, $d(I_{s'})= 6$
          and $d(I_s  \cap  I_{s'}) = 6$ if $s = s'$ and $d(I_s  \cap  I_{s'})= 4$ if $s \neq
          s'$ (due to intersection rules (\ref{2.8})). Hence we
          get from (\ref{A.16}), (\ref{A.17}) and (\ref{A.18}):
         ${\cal M}_{s}^{s'} = - 1$ for $s = s'$ and
          ${\cal M}_{s}^{s'} =  0$ for $s \neq s'$
         in agreement with (\ref{A.19}). Thus,  relation (\ref{A.19})
         is proved and hence  relations (\ref{A.10}) and (\ref{2.15b})
         are also proved.

          \subsection{The proof of the Proposition 1}

   Relations (\ref{0.36}) and (\ref{0.37}) may be written in a
   condensed form as follows
   \beq{A.3}
    D_{m_l} = \bar{D}_{m_l}^{(l)} +  \sum_{s \in S} A^s_{m_l},
   \eeq
   $l = 0,1, \dots, n$, ($m_0 = \mu$) where
   \beq{A.4}
    A^s_{m_l} =  \frac{1}{4}  \delta_0^l
    (\Gamma_{m_0} \Gamma^{\nu} - \Gamma^{\nu} \Gamma_{m_0} ) \gamma_{s,
   \nu}
   + \frac{1}{2} (1 - \delta_0^l)  \Gamma_{m_l} \Gamma^{\nu} \phi^{l}_{s, \nu},
   \eeq
   with $\gamma_s$ and  $\phi^i_s$ defined in (\ref{A.7}) and
   (\ref{A.8}).

      Due to ansatz  (\ref{2.16}) and (\ref{A.3}) the SUSY
   equations (\ref{1.2}) read
   \beq{A.5}
   [ \bar{D}_{m_l}^{(l)}  +  \sum_{s \in S}
      ( A^s_{m_l} + B^s_{m_l} + b_s \p_{m_l} \ln H_s ) ] \eta = 0,
   \eeq
    $l = 0,1, \dots, n$, where $b_s = - 1/6$ if $s \in S_e$ and
    $b_s = - 1/12$ if $s \in S_m$.

    Relation (\ref{A.5}) is valid due to
    condition (\ref{2.16a}) ($\bar{D}_{m_l}^{(l)} \eta = 0$ )
    and the following identities

    \beq{A.6}
    ( A^s_{m_l} + B^s_{m_l} + b_s \p_{m_l} \ln H_s ) \eta = 0,
    \eeq
    $l = 0,1, \dots, n$, $s \in S$, which are valid when the chirality
    restrictions (\ref{2.16b}) are imposed. The verification of
    identities (\ref{A.6}) is a straightforward one for electric and magnetic
    branes by using formulae (\ref{2.10}) and (\ref{2.11}).

      \subsection{Example: chiral parallel spinors on product of two 4-dimensional manifolds}

     Let us consider chiral parallel spinors on $M_0 \times M_1$,
     where $M_0$ and $M_1$ are 4-dimensional  Ricci-flat manifolds of
     Euclidean signatures.

  We consider $\Gamma$-matrices
  \beq{A.3n}
   (\hat{\Gamma}^A) = (\hat{\Gamma}^{a_0}_{(0)} \otimes
  \1_4, \hat{\Gamma}_{(0)} \otimes \hat{\Gamma}^{a_1}_{(1)}),
  \eeq
 where $\hat{\Gamma}^{a_0}_{(0)}$, $a_0 =
 1_0,  \ldots, 4_0$, correspond to $M_0$,
 and    $\hat{\Gamma}^{a_1}_{(1)}$,   $a_1 = 1_1, ..., 4_1$,
 correspond to $M_1$. Here  $\hat{\Gamma}_{(0)}
  =   \hat{\Gamma}^{1_0}_{(0)} \hat{\Gamma}^{2_0}_{(0)} \hat{\Gamma}^{3_0}_{(0)}  \hat{\Gamma}^{4_0}_{(0)}$
 obeys $(\hat{\Gamma}_{(0)})^2 =  \1_{4}$.

 We obtain
     \bear{A.6a}
     \bar{D}_{m_0}^{(0)} = \p_{m_0} + \frac{1}{4}
     \omega^{(0)}_{a_0 b_0 m_0}  \hat{\Gamma}^{a_0}
     \hat{\Gamma}^{b_0} \otimes \1_4, \\ \label{A.6b}
     \bar{D}_{m_1}^{(1)} = \p_{m_1} + \frac{1}{4}
     \omega^{(1)}_{a_1 b_1 m_1} \1_{4} \otimes \hat{\Gamma}^{a_1}
     \hat{\Gamma}^{b_1}.
   \ear

   Let
   \beq{A.6c}
   \eta =  \eta_0(x) \otimes \eta_1(y_1),
   \eeq
    where $\eta_0  = \eta_0(x)$ is $4$-component spinor on $M_0$
  and   $\eta_1 =  \eta_1(y_1)$ is $4$-component spinor on
   $M_1$. Then
    \beq{A.7a}
     \bar{D}_{m_0}^{(0)} \eta  =  (D_{m_0}^{(0)} \eta_0) \otimes \eta_1,
     \qquad
     \bar{D}_{m_1}^{(1)} \eta =    \eta_0  \otimes (D_{m_1}^{(1)}
     \eta_1),
     \eeq
  where $D^{(0)}_{m_0} = \p_{m_0} +  \frac{1}{4} \omega^{(0)}_{a_0 b_0 m_0}
  \hat{\Gamma}^{a_0}_{(0)} \hat{\Gamma}^{b_0}_{(0)}$ and $D^{(1)}_{m_1} =
  \p_{m_1} +  \frac{1}{4} \omega^{(1)}_{a_1 b_1 m_1}
   \hat{\Gamma}^{a_1}_{(1)} \hat{\Gamma}^{b_1}_{(1)}$
   are covariant (spinorial) derivatives  corresponding to the manifolds
   $M_0$ and $M_1$, respectively.

   It follows from the Proposition 2 and (\ref{A.7a}) that $\eta =  \eta_0(x) \otimes
   \eta_1(y_1)$ is parallel spinor if and only if $\eta_0$ and  $\eta_1$
   are parallel spinors on $M_0$ and $M_1$, respectively.

  The  chirality operator $\hat{\Gamma}$ (which is a product of all
  $\hat{\Gamma}^A$) reads

   \beq{A.8a}
     \hat{\Gamma} = \hat{\Gamma}_{(0)} \otimes \hat{\Gamma}_{(1)},
     \eeq
  where  $\hat{\Gamma}_{(1)} = \hat{\Gamma}^{1_1}_{(1)}
  \hat{\Gamma}^{2_1}_{(1)} \hat{\Gamma}^{3_1}_{(1)} \hat{\Gamma}^{4_1}_{(1)}$
obeys $(\hat{\Gamma}_{(1)})^2 =  \1_{4}$.

If $\eta_i$ is chiral parallel spinor on $M_i$ with chirality
$c_{(i)} = \pm 1$: $\hat{\Gamma}_{(i)} \eta_i = c_{(i)} \eta_i$,
$i = 0, 1$, then $\eta =  \eta_0 \otimes \eta_1$ is chiral
parallel spinor on $M_0 \times M_1$, with the chirality $c =
c_{(0)} c_{(1)}$: $\hat{\Gamma} \eta = c \eta$.

The generalization of this example to arbitrary even dimensions
 $d_0$ and $d_1$ is a straightforward one.


  \end{document}